\documentclass[
  twocolumn,
  prb,
  showpacs,
  amsmath,
  amssymb,
  superscriptaddress,
  floatfix
]{revtex4}

\usepackage{bm}
\usepackage{graphicx}
\usepackage{color}

\DeclareMathOperator{\Tr}{Tr}
\DeclareMathOperator{\spp}{sp}
\DeclareMathOperator{\sgn}{sgn}
\DeclareMathOperator{\liq}{li_2}
\DeclareMathOperator{\lit}{li_4}
\DeclareMathOperator{\im}{Im}
\DeclareMathOperator{\re}{Re}

\begin{document}

\title{Multifractality and electron-electron interaction at Anderson transitions}

\author{I.S.~Burmistrov}

\affiliation{ L.D. Landau Institute for Theoretical Physics, Kosygina
  street 2, 117940 Moscow, Russia}
  
\affiliation{Moscow
Institute of Physics and Technology, 141700 Moscow, Russia}

\author{I.V.~Gornyi}
\affiliation{
 Institut f\"ur Nanotechnologie, Karlsruhe Institute of Technology,
 76021 Karlsruhe, Germany
}
\affiliation{
 A.F.~Ioffe Physico-Technical Institute,
 194021 St.~Petersburg, Russia.
}

\author{A.D.~Mirlin}
\affiliation{
 Institut f\"ur Nanotechnologie, Karlsruhe Institute of Technology,
 76021 Karlsruhe, Germany
}
\affiliation{
 DFG Center for Functional Nanostructures,
 Karlsruhe Institute of Technology, 76128 Karlsruhe, Germany
}
\affiliation{
 Inst. f\"ur Theorie der kondensierten Materie,
 Karlsruhe Institute of Technology, 76128 Karlsruhe, Germany
}
\affiliation{
 Petersburg Nuclear Physics Institute,
 188300 St.~Petersburg, Russia.
}

\begin{abstract}
Mesoscopic fluctuations and correlations of
the local density of states are studied 
near metal-insulator transitions in disordered interacting electronic systems. We show that the multifractal behavior of the local density of states survives in the presence of Coulomb
interaction. We calculate the spectrum of multifractal exponents in 
$2+\epsilon$ spatial dimensions for symmetry classes characterized by broken (partially or fully) spin-rotation invariance  
and show that it differs from that in the absence of
interaction. We also estimate the multifractal exponents at the Anderson metal-insulator transition in 2D systems with preserved spin-rotation invariance.   Our results for multifractal correlations of the local density of states are in qualitative agreement with recent experimental findings.

\end{abstract}


\maketitle

\section{Introduction}

Anderson localization and Anderson localization transitions
remain an actively developing field [\onlinecite{AL50,Evers08}]. 
Metal-insulator transitions and quantum Hall plateau transitions have been experimentally observed
and studied in a variety of semiconductor structures [\onlinecite{semicond}]. Recent discoveries of graphene [\onlinecite{graphene}] and time-reversal-invariant topological insulator materials [\onlinecite{topins}] considerably extended the scope of experimental and theoretical research on Anderson localization and its interplay with topology.

Similarly to other quantum phase transitions, Anderson transitions  are characterised by critical scaling of various physical observables. Remarkably, wave  functions at Anderson transitions demonstrate multifractal behavior which imply 
their very strong fluctuations. Specifically, the wave-function moments
(the averaged inverse participation ratios)
\begin{equation}
\langle P_q\rangle=\left \langle\,\, \int\limits_{r<L} d^d\bm{r} \bigl |\psi(\bm{r})\bigr |^{2q}\right \rangle  .
\end{equation}
show at the transition point an anomalous multifractal scaling with respect to the
system size $L$,
\begin{equation}\label{e1}
 L^d \langle |\psi(\bm{r})|^{2q} \rangle \propto L^{-\tau_q}, \qquad \tau_q =
d(q-1) +  \Delta_q .
\end{equation}
Here $d$ is the spatial dimension and $\langle\dots\rangle$ denotes the
averaging over disorder. The anomalous multifractal
exponents  $\Delta_q$ are negative at the critical point whereas in a conventional metallic
phase one finds $\Delta_q\equiv 0$. The result \eqref{e1} implies the following scaling for the moments of the local density of states at the critical point:
\begin{equation}\label{e1-rho}
 \langle [\rho(E,\bm{r})^{q} ] \rangle \propto L^{-\Delta_q} .
\end{equation}
In fact, the nontrivial behavior of wave functions at Anderson transitions 
is much reacher than only scaling of averaged participation ratios  {[\onlinecite{Wegner}]}. Recently, a
complete classification of observables characterizing critical wave functions
(that includes multifractal moments (\ref{e1}) as a ``tip of the iceberg'')
was developed [\onlinecite{gruzberg13}].

The above results on multifractality have been obtained for
disordered systems in the absence of the electron-electron interaction. In most cases, even a short-ranged interaction is relevant perturbation in the renormalization-group (RG) sense. The only exception when noninteracting multifractality remains valid in the presence of interaction is the case of broken time reversal symmetry and broken spin invariance for which the short-range (e.g., screened by external gate) electron-electron interaction is RG-irrelevant.  {In this case the scaling behavior of critical wave functions determines the interaction-induced dephasing at the non-interacting critical point [\onlinecite{HW,WFGC,BBEGM}].}
For systems of other symmetries, the short-range electron-electron interaction is RG-relevant. However, the multifractality of non-interacting electrons has a
dramatic impact on properties of such systems, as  it determines the RG evolution of the system away
from the non-interacting fixed point. In particular, it was shown that the
multifractality leads to a strong enhancement of superconducting transition
temperature [\onlinecite{feigelman07,burmistrov12,dellanna}]  and  controls possible instabilities of
surface states of topological superconductors with respect to interaction
[\onlinecite{foster12,foster14}].

The long-range ($1/r$) Coulomb interaction
is always RG-relevant at the non-interacting fixed point 
and thus has a strong impact on localization properties of the
system (see Refs. [\onlinecite{finkelstein90,belitz94,finkelstein10}] for review). In view of a combination of disorder and interaction physics, metal-insulator transitions in the presence of Coulomb interaction are often called Mott-Anderson (or
Anderson-Mott) transitions.

One of distinctive manifestations of the Coulomb interaction is a strong suppression of the local density of states  
at the Fermi level [\onlinecite{AA1979AAL1980,ES75SE84}]. The
local density of states can be measured in a tunneling experiment, implying a suppression of the tunneling  current at low bias voltages, which is known as zero-bias anomaly. Specifically, in a two-dimensional (2D) weakly disordered
system the disorder-averaged local density of states behaves as
[\onlinecite{Fin198384,Castellani1984,nazarov89,levitov97,kamenev99}]
\begin{equation}
\langle \rho(E) \rangle \propto \exp\left\{-{1\over 4\pi g} \ln^2 |E|
\right\} ,
\label{e2}
\end{equation}
where $g$ is the dimensionless (measured in units $e^2/h$) conductivity and energy $E$ is counted from the chemical potential. 
The unconventional behavior \eqref{e2} with squared
logarithm in the exponential (rather than with a simple logarithm that would
yield a power law, as normally expected for critical behavior) is due to gauge-type phase fluctuations which leads to a suppression of Debye-Waller type. For the Anderson transition in
$d=2+\epsilon$ dimension (with $\epsilon \ll 1$ allowing a parametric control
of the theory) the disorder-averaged local density of states demonstrates the following scaling behavior
\begin{equation}
\langle \rho(E) \rangle \propto |E|^\beta, \qquad \beta = O(1),
\label{e3}
\end{equation}
with precise value of the critical exponent $\beta$ depending on the
symmetry class. Equation (\ref{e3}) originates from Eq.~(\ref{e2}) when one passes from two to $2+\epsilon$ dimensions due to a transformation of  one of the logarithmic factors in Eq.~(\ref{e2}) into a factor $\sim
1/\epsilon$ and because the critical conductance  
$g_*$ is of order $1/\epsilon$ [\onlinecite{finkelstein90,belitz94}].  
We remind that in the absence of interaction the average local density of states is uncritical, $\beta=0$, 
in conventional (Wigner-Dyson) symmetry classes.

We are thus facing the following important question: Does multifractality survive in the presence of Coulomb interaction between electrons? This question is of direct experimental relevance. The scanning tunneling microscopy (STM) provides a direct access to the fluctuations and correlations of the local density of states in disordered interacting systems. In particular, multifractal fluctuations and correlations of the local density of states have been measured recently [\onlinecite{richardella10}] in a magnetic semiconductor Ga$_{1-x}$Mn$_x$As near metal-insulator transition. Strong fluctuations of the local density of states in a strongly disordered 3D system (presumably, on the insulating side of the transition) have been observed in Ref.~[\onlinecite{morgenstern02}]. Recent STM measurement in various 2D semiconductor systems and graphene [\onlinecite{morgenstern-2D}] demonstrated the feasibility to explore
fluctuations and correlations of the local density of states also near the quantum Hall
transitions. Strong spatial fluctuations of the  local density of states have been also
reported near the superconductor-insulator transition in disordered films [\onlinecite{sacepe08}]. 

The experimental findings mentioned above suggest that the multifractality of the local density of states does survive in the presence of Coulomb interaction. Recently, this conclusion was supported by numerical analysis in the framework of the density functional theory [\onlinecite{slevin}], by the Hartree-Fock simulation of the problem [\onlinecite{kravtsov14}], and by the authors within the nonlinear sigma model analysis in $d=2+\epsilon$ dimension in the case of broken time reversal and spin rotational symmetries [\onlinecite{BGM2013}].

In this paper, we extend the nonlinear sigma model analysis of Ref. [\onlinecite{BGM2013}] to all conventional symmetry cases. 
Specifically, we consider models with fully broken, partially broken, and preserved spin rotation invariance, both with and without time-reversal symmetry.  We will demonstrate that in spite of the suppression of the disorder average $\langle\rho(E)\rangle$ due to gauge-type phase fluctuations, the normalized local density of states $\rho(E,\bm{r})/\langle\rho(E)\rangle$
in strongly interacting critical systems does show multifractal
fluctuations and correlations. We will also calculate the spectrum
of anomalous dimensions in $2+\epsilon$ spatial dimensions in the
two-loop approximation (up to the $\epsilon^2$ order) for symmetry cases with broken (partially or fully) spin-rotation invariance and demonstrate that it differs from that of corresponding non-interacting symmetry classes. In the case of preserved spin-rotation invariance, our results yield an estimate for the multifractal exponents at the 2D metal-insulator transition  {in the model with large number of valleys, $n_v \gg 1$, studied in Ref. [\onlinecite{punnoose05}]}. 

The outline of the paper is as follows. In Sec. \ref{s2} we introduce formalism of the nonlinear sigma model. Details of the two-loop renormalization group analysis for moments of the local density of states in $d=2+\epsilon$
dimension are given in Sec. \ref{s3}. In Sec. \ref{s4.1} we present results of a general scaling analysis for moments as well as spatial and energy correlations of the local density of states near the interacting critical point. The results for a set of multifractal exponents for  various symmetry classes are presented in Secs. \ref{s4.2} - \ref{s4.5}. We conclude the paper with the summary of our findings and a discussion of open questions, Sec.~\ref{s5}. Some of the results were published in a brief form in Ref. [\onlinecite{BGM2013}].

\section{Formalism}

\label{s2}

\subsection{Nonlinear sigma model action}

For the case of preserved time reversal and spin rotational symmetries the action of the  nonlinear sigma model (NLSM) is given as the sum of the non-interacting part, $S_\sigma$,
and contributions arising from the interactions in the particle-hole singlet and triplet channels, and the particle-particle (Cooper) channel, $S_{\rm int}$: [\onlinecite{finkelstein90,belitz94}]
\begin{gather}
S=S_\sigma + S_{\rm int},
\label{eq:NLSM}
\end{gather}
where
\begin{equation}
S_\sigma  = -\frac{g}{32} \int d\bm{r} \Tr (\nabla Q)^2 + 4\pi T z \int d\bm{r} \Tr \eta Q .
\label{Ss}
\end{equation}
 {In this paper we focus on systems with repulsive interaction in the Cooper channel, i.e., systems without a superconducting instability. Thus we neglect the interaction in the Cooper channel. Then, the interaction part of the action can be written as}  
\begin{align}
S_{\rm int} & =- \frac{\pi T}{4} \sum_{\alpha,n} \sum_{r=0,3}\sum_{j=0}^3\Gamma_j 
\int d\bm{r} \Tr \Bigl [I_n^\alpha t_{rj} Q\Bigr ] \notag  \\
& \hspace{3cm}\times \Tr \Bigl [I_{-n}^\alpha t_{rj} Q\Bigr ] .
\label{Srho}
\end{align}
Here $g$ is the total Drude conductivity
(in units $e^2/h$ and including spin), $\Gamma_0 = \Gamma_s$  denotes the interaction amplitude in the singlet channel, and  $\Gamma_1=\Gamma_2=\Gamma_3=\Gamma_t$ stands for the interaction amplitude in the triplet channel. The parameter $z$ is frequency renormalization factor introduced by Finkelstein [\onlinecite{Fin198384}]. We use the following matrices
\begin{gather}
\Lambda_{nm}^{\alpha\beta} = \sgn n \, \delta_{nm} \delta^{\alpha\beta}t_{00}, \quad
\eta_{nm}^{\alpha\beta}=n \, \delta_{nm}\delta^{\alpha\beta} t_{00},  \notag \\
(I_k^\gamma)_{nm}^{\alpha\beta}=\delta_{n-m,k}\delta^{\alpha\beta}\delta^{\alpha\gamma} t_{00} , 
\end{gather}
with $\alpha,\beta = 1,\dots, N_r$ standing for replica indices and $n,m$ corresponding to the
Matsubara fermionic energies $\varepsilon_n = \pi T (2n+1)$. The sixteen matrices,
\begin{equation}
\label{trj}
t_{rj} = \tau_r\otimes s_j, \qquad r,j = 0,1,2,3  ,
\end{equation}
operate in the particle-hole (subscript $r$) and spin (subscript $j$) spaces
with the corresponding Pauli matrices denoted by
\begin{equation}
\begin{split}
\tau_1 & = \begin{pmatrix}
0 & 1\\
1 & 0
\end{pmatrix}, \, \tau_2 = \begin{pmatrix}
0 & -i\\
i & 0
\end{pmatrix}, \, \tau_3 = \begin{pmatrix}
1 & 0\\
0 & -1
\end{pmatrix} , \label{eq:tau-def}\\
s_1 & = \begin{pmatrix}
0 & 1\\
1 & 0
\end{pmatrix}, \, s_2 = \begin{pmatrix}
0 & -i\\
i & 0
\end{pmatrix}, \, s_3 = \begin{pmatrix}
1 & 0\\
0 & -1
\end{pmatrix} .
 \end{split}
\end{equation}
Matrices $\tau_0$ and $s_0$ stand for the $2\times 2$ unit matrices. The matrix field $Q(\bm{r})$ (as well as the trace $\Tr$) acts in the replica, Matsubara,
spin, and particle-hole spaces. It obeys the following constraints:
\begin{gather}
Q^2=1, \qquad \Tr Q = 0, \qquad Q^\dag = C^T Q^T C .
\end{gather}
 The charge conjugation matrix $C = i t_{12}$ satisfies the following relation $C^T = -C$.
Matrix $Q$ can be parameterized as $Q=T^{-1} \Lambda T$ where the matrices $T$ obey
 \begin{equation}
 C T^* = T C,\qquad (T^{-1})^* C = C T^{-1} . \label{TC}
 \end{equation}
Symbol $*$ denotes the complex conjugation.

\subsection{Moments of the local density of states}
 
The local density of states is determined by the single-particle Green function as follows: $\rho(E, \bm{r}) = (-1/\pi)\im G(E; \bm{r}, \bm{r})$. Within NLSM formalism the disorder-averaged local density of states $\langle \rho(E) \rangle$ can be obtained after analytic continuation to the real energies, $i\varepsilon_n \to E+i0^+$, of the following correlation function
\begin{equation}
\rho(i\varepsilon_n) = \frac{\rho_0}{4} \spp \langle Q_{nn}^{\alpha\alpha}\rangle .
\label{eq:PO:TDOS}
\end{equation}
Here $\alpha$ is a fixed replica index, symbol $\spp$ denotes trace in spin and particle-hole spaces, and $\rho_0$ stands for the single-particle density of states (including spin) at energy of the order of inverse elastic scattering time $1/\tau$ which plays a role of high-energy (ultra-violet) cutoff of the theory. 
 
In order to discuss the second moment of the local density of states it is convenient to introduce
the irreducible two-point correlation function
\begin{gather}
K_2(E,\bm{r};E^\prime, \bm{r^\prime}) = \langle \langle \rho(E, \bm{r})  \cdot \rho(E^\prime, \bm{r^\prime}) \rangle \rangle \hspace{2cm}\notag \\
\hspace{1cm} =  \langle \rho(E, \bm{r}) \rho(E^\prime, \bm{r^\prime}) \rangle  - \langle \rho(E, \bm{r}) \rangle \langle \rho(E^\prime, \bm{r^\prime}) \rangle .
\label{eq:K2:def:gen}
\end{gather}
At coinciding spatial points this function, $K_2(E,\bm{r};E^\prime, \bm{r})$, can be obtain from
 \begin{equation}
 K_2 = \frac{\rho_0^2}{32}\re \Bigl [ P_2^{\alpha_1\alpha_2}(i\varepsilon_{n_1},i\varepsilon_{n_3}) - P_2^{\alpha_1\alpha_2}(i\varepsilon_{n_1},i\varepsilon_{n_2}) \Bigl ] 
 \label{eqK2def0}
 \end{equation}
after analytic continuation to the real frequencies:  $\varepsilon_{n_1} \to E+i0^+$, $\varepsilon_{n_3} \to E^\prime+i0^+$, and $\varepsilon_{n_2} \to E^\prime-i0^+$. Here 
 \begin{gather}
 P_2^{\alpha_1\alpha_2}(i\varepsilon_{n},i\varepsilon_{m}) = \langle\langle  \spp Q_{nn}^{\alpha_1\alpha_1}(\bm{r}) \cdot \spp Q_{mm}^{\alpha_2\alpha_2}(\bm{r}) \rangle \rangle \notag  \\ - 2 \langle \spp \bigl [Q_{nm}^{\alpha_1\alpha_2}(\bm{r}) Q_{mn}^{\alpha_2\alpha_1}(\bm{r}) \bigr ] \rangle 
  .
 \label{eqP2corr}
 \end{gather}
Replica indices $\alpha_1$ and $\alpha_2$ are different in Eq. \eqref{eqK2def0}, $\alpha_1\neq \alpha_2$, so that the two-point correlation function $K_2$ measures mesoscopic fluctuations of the local density of states. 
 {We mention that the NLSM operator corresponding to $K_2$ is the eigenoperator under action of the renormalization group as we shall explicitly check by two-loop calculations below.} In a similar way, higher moments of the local density of states and corresponding irreducible correlation functions can be expressed in terms of higher-order correlation functions of the $Q$-field  {which are eigenoperators of the renormalization group}.

\section{Two-loop renormalization}
\label{s3}

\subsection{Perturbative expansion}

For the perturbative treatment (in $1/g$) of the NLSM action \eqref{eq:NLSM}
we shall use the square-root parametrization
\begin{gather}
Q = W +\Lambda \sqrt{1-W^2}\ , \qquad W= \begin{pmatrix}
0 & w\\
\bar{w} & 0
\end{pmatrix} .
\label{eq:Q-W}
\end{gather}
We adopt the following notations: $W_{n_1n_2} = w_{n_1n_2}$ and $W_{n_2n_1} = \bar{w}_{n_2n_1}$ with $n_1\geqslant 0$ and $n_2< 0$.
The blocks $w$ and $\bar{w}$ (in Matsubara space) obey
\begin{gather}
\bar{w} = -C w^T C,\qquad w = - C w^* C .
\end{gather}
The second equality implies that in the expansion $w^{\alpha\beta}_{n_1n_2}= \sum_{rj} (w^{\alpha\beta}_{n_1n_2})_{rj} t_{rj}$ some of the elements
$(w^{\alpha\beta}_{n_1n_2})_{rj}$ are real and some are purely imaginary.

Expanding the NLSM action \eqref{eq:NLSM} to the second order in $W$, we find the following propagators for diffusive modes. The propagators of diffusons read ($r=0,3$ and $j=0,1,2,3$)
\begin{gather}
\Bigl \langle [w_{rj}(\bm{p})]^{\alpha_1\beta_1}_{n_1n_2} [\bar{w}_{rj}(-\bm{p})]^{\beta_2\alpha_2}_{n_4n_3} \Bigr \rangle =  \frac{2}{g} \delta^{\alpha_1\alpha_2} \delta^{\beta_1\beta_2}\delta_{n_{12},n_{34}}\notag \\
\times  \mathcal{D}_p(i\Omega_{12}^\varepsilon)\Bigl [\delta_{n_1n_3} - \frac{32 \pi T \Gamma_j}{g}\delta^{\alpha_1\beta_1}  \mathcal{D}_p^{(j)}(i\Omega_{12}^\varepsilon) \Bigr ] ,
\label{eq:prop:PH}
\end{gather}
where $\Omega_{12}^\varepsilon = \varepsilon_{n_1}-\varepsilon_{n_2}$.
The standard diffusive propagator is given as
\begin{equation}
\mathcal{D}^{-1}_p(i\omega_n) =p^2+{16 z |\omega_n|}/{g} .
\label{eq:prop:Free}
\end{equation}
The diffusons renormalized by interaction in the singlet  ($\mathcal{D}_p^{(0)}(\omega) \equiv \mathcal{D}_p^{s}(\omega)$) and triplet  ($\mathcal{D}_p^{(1)}(\omega)=\mathcal{D}_p^{(2)}(\omega)=\mathcal{D}_p^{(3)}(\omega) \equiv \mathcal{D}_p^{t}(\omega)$)   particle-hole channels are as follows
\begin{align}
[\mathcal{D}^s_p(i\omega_n)]^{-1} & =  p^2+{16 (z+\Gamma_s) |\omega_n|}/{g},\notag \\
 [\mathcal{D}^t_p(i\omega_n)]^{-1} & =  p^2+{16 (z+\Gamma_t) |\omega_n|}/{g} .
 \label{eq:prop:Int}
\end{align}
The propagators of singlet and triplet cooperons ($r=1,2$ and $j=0,1,2,3$) are insensitive to the interaction in the particle-hole channels:
\begin{align}
\Bigl \langle [w_{rj}(\bm{p})]^{\alpha_1\beta_1}_{n_1n_2} [\bar{w}_{rj}(-\bm{p})]^{\beta_2\alpha_2}_{n_4n_3} \Bigr \rangle & =  \frac{2}{g} \delta^{\alpha_1\alpha_2} \delta^{\beta_1\beta_2} \delta_{n_1n_3}
\notag \\
&\times
 \delta_{n_2n_4}\mathcal{C}_p(i\Omega_{12}^\varepsilon) ,
 \label{eq:prop:PPT}
\end{align}
where $\mathcal{C}_p(i\omega_n) \equiv \mathcal{D}_p(i\omega_n)$.

For the purpose of regularization in the infrared, it is convenient to add the following term to the NLSM action 
\eqref{eq:NLSM}:
\begin{equation}
S \to S + \frac{g h^2}{8} \int d \bm{r}\Tr \Lambda Q .
\label{SsGenFull}
\end{equation}
This leads to the substitution of $p^2 +h^2$ for $p^2$ in the propagators \eqref{eq:prop:Free} and \eqref{eq:prop:Int}.

 \subsection{The disorder-averaged local density of states}

We start from renormalization of the disorder-averaged local density of states. For our purposes, it is enough to compute it in the one-loop approximation. Expanding the matrix $Q$ to the second order in $W$ and using Eq. \eqref{eq:PO:TDOS}, we obtain
\begin{equation}
\frac{\rho(i\varepsilon_{n_1})}{\rho_0}  = 1 - \frac{1}{8} \spp \sum_{n_2,\beta}  \langle w_{n_1n_2}^{\alpha\beta}(\bm{r}) \bar{w}_{n_2n_1}^{\beta\alpha}(\bm{r})  \rangle  .
\end{equation}
Computing the average with the help of Eqs. \eqref{eq:prop:PH} - \eqref{eq:prop:PPT} we find
\begin{gather}
\frac{\rho(i\varepsilon_{n_1})}{\rho_0}    = 1 + \frac{64\pi T}{g^2} \int_q \sum_{\omega_n>\varepsilon_{n_1}}  \sum_{j=0}^3 \Gamma_j \mathcal{D}_q(i\omega_n) \mathcal{D}^{(j)}_q(i\omega_n)
.
\label{app:eq_P1}
\end{gather}
Finally,  performing analytic continuation to the real frequencies, $i \epsilon_{n_1} \to E + i 0^+$, we obtain
\begin{align}
\frac{\rho(E)}{\rho_0}  = 1  & +  \frac{16}{g^2} \im  \sum_{j=0}^3 \Gamma_j
\int_{q,\omega}  \mathcal{F}_{\omega-E}   \mathcal{D}^R_q(\omega) \mathcal{D}^{(j) R}_q(\omega)
 .
\label{rhores}
\end{align}
Here $\mathcal{D}^R_q(\omega)$ and $\mathcal{D}^{(j) R}_q(\omega)$ are
retarded propagators corresponding to Matsubara propagators $\mathcal{D}^R_q(i\omega_n)$ and $\mathcal{D}^{(j) R}_q(i\omega_n)$, respectively. The fermionic distribution function is denoted as $\mathcal{F}_\omega = \tanh(\omega/2T)$. We use the following short-hand notation:
\begin{equation}
\int_{q,\omega} \equiv \int \frac{d^d\bm{q}}{(2\pi)^d} \int\limits_{-\infty}^\infty d\omega  .
\end{equation}
Since $\Gamma_s \mathcal{D}^R_q(\omega)\mathcal{D}_q^{s R}(\omega) \propto [\mathcal{D}^{R}_q(\omega)]^2U_{\rm{scr}}(\omega,\bm{q})$
 where $U_{\rm{scr}}(\omega,\bm{q})$ is dynamically screened Coulomb interaction, one can check that Eq.~\eqref{rhores} reproduces the well-known perturbative result for the zero-bias anomaly [\onlinecite{AA1979AAL1980}].

The result \eqref{rhores} implies that the disorder-averaged local density of states can be written as $\rho(E) = \rho_0 [Z(E)]^{1/2}$ with the renormalization factor 
\begin{gather}
Z(E) = 1  +  \frac{16}{g^2}  \sum_{j=0}^3  \Gamma_j \int_{q,\omega}  \bigl [\mathcal{F}_{\omega-E}+\mathcal{F}_{\omega+E} \bigr ] \notag \\
\times \im \bigl [ \mathcal{D}^R_q(\omega) \mathcal{D}^{(j) R}_q(\omega) \bigr ].
\label{eq_Z}
\end{gather}
 {We note that such definition of $Z$ coincides with the definition of the field renormalization constant in Ref. [\onlinecite{belitz94}] and the wave-function renormalization constant in Ref. [\onlinecite{Castellani1984}]. We stress that $Z$ is very different from the frequency renormalization factor $z$ introduced by Finkelstein [\onlinecite{Fin198384}]. }

To simplify analysis, it is convenient to set temperature $T$ and energy $E$ to zero and study dependence of $\langle \rho\rangle$ on the infrared regulator $h^2$. 
Hence, in $d=2+\epsilon$ dimension, we obtain [\onlinecite{Castellani1984}]
 \begin{equation}
Z = 1 - \bigl [\ln(1+\gamma_s)+3\ln(1+\gamma_t)\bigr ]\frac{h^\epsilon t}{\epsilon} +O(\epsilon) .
\label{eqAvDOS}
\end{equation}
Here $\gamma_s=\Gamma_s/z$ and $\gamma_t=\Gamma_t/z$ are dimensionless interaction amplitudes and $t = 8 \Omega_d/g$ denotes resistivity, where $\Omega_d= S_d/[2(2\pi)^d]$ and $S_d=2\pi^{d/2}/\Gamma(d/2)$ is the area of the $d$-dimensional sphere. We notice the well-known peculiarity of the case of Coulomb interaction ($\gamma_s=-1$) for which the formally divergent term $ \ln(1+\gamma_s)$ in Eq. \eqref{eqAvDOS} emerges in addition to $1/\epsilon$ factor. 

\begin{figure}[t]
\centerline{\includegraphics[width=40mm]{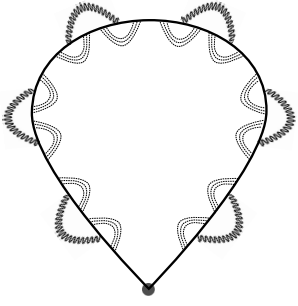}}
\caption{Representative diagram for the disorder-averaged local density of states. Solid lines denote electron Green functions, while wavy lines denote the dynamically screened Coulomb interaction. Ladders of dashed lines dressing the interaction vertices represent diffusons. }
\label{Figure1rho}
\end{figure}

As usual, Eq. \eqref{eqAvDOS} determines the anomalous dimension $\zeta$ of the disorder-averaged local density of states. In the one-loop approximation, we obtain
\begin{equation}
- \frac{d \ln Z}{d y} = 2 \zeta = - \bigl [\ln(1+\gamma_s)+3\ln(1+\gamma_t)\bigr ] t + O(t^2) , 
\label{eq:ad:Z}
\end{equation} 
where $y=-\ln 1/h$ running renormalization group length scale. 
To illustrate the renormalization group result \eqref{eq:ad:Z}, we show in Fig. \ref{Figure1rho} a representative
diagram for the disorder-averaged local density of states. The local density of states is given by a fermionic loop dressed by interaction lines. Averaging the loop over disorder generates diffusive vertex corrections and yields the suppression of the average local density of states by gauge-type phase fluctuations.

 \subsection{The second moment of the local density of states}
 
 The renormalization of the second moment of the local density of states can be extracted from the irreducible two-point correlation function $K_2$ at the coinciding spatial points. We limit our consideration by one- and two-loop orders. 
 
 \subsubsection{One-loop results}
 
In the one-loop approximation we find
 \begin{gather}
[P_2^{\alpha_1\alpha_2}]^{(1)}(i\varepsilon_{n_1},i\varepsilon_{n_3}) = 0, 
\end{gather}
and
\begin{gather}
[P_2^{\alpha_1\alpha_2}]^{(1)}(i\varepsilon_{n_1},i\varepsilon_{n_2}) = - 2 \spp \langle w^{\alpha_1\alpha_2}_{n_1n_2}(\bm{r}) \bar{w}^{\alpha_2\alpha_1}_{n_2n_1}(\bm{r}) \rangle \notag \\ = - \frac{128}{g} \int_q 
\Bigl [ \mathcal{D}_q(i\Omega^{\varepsilon}_{12})+\mathcal{C}_q(i\Omega^{\varepsilon}_{12})\Bigr] 
.  \label{eq18P1}
\end{gather}
Hence,
we obtain
\begin{equation}
K_2^{(1)}(E,\bm{r};E^\prime, \bm{r})  = \rho_0^2\,\frac{4 }{g} \re \int_q \Bigl [  \mathcal{D}^R_q(\Omega) + 
 \mathcal{C}^R_q(\Omega)\Bigr ]  , \label{eq1loopK2_2}
\end{equation}
where $\Omega=E-E^\prime$. Setting $E=E^\prime$ and using $h^2$ as the infrared regulator, one finds
\begin{equation}
K_2^{(1)}  =  - \rho_0^2 \frac{2 h^\epsilon t}{\epsilon} +O(\epsilon) . \label{eq1loopK2_3}
\end{equation}

 \subsubsection{Two-loop results}
 
 We start evaluation of the two-loop contribution to the irreducible two-point correlation function $K_2$ from $P^{\alpha_1\alpha_2}_2(i\varepsilon_{n_1},i\varepsilon_{n_3})$. In the two-loop approximation, one needs to take into account only terms with four $W$: 
\begin{gather}
[P_2^{\alpha_1\alpha_2}]^{(2)}(i\varepsilon_{n_1},i\varepsilon_{n_3}) = \frac{1}{4} \sum_{n_6n_8}\sum_{\beta_1\beta_2} \Bigl [ \langle \langle \spp \bigl [
w^{\alpha_1\beta_1}_{n_1n_6}\bar{w}^{\beta_1\alpha_1}_{n_6n_1} \bigr ] \notag \\
\cdot \spp \bigl [
w^{\alpha_2\beta_2}_{n_3n_8}\bar{w}^{\beta_2\alpha_2}_{n_8n_3}  \bigr ]
\rangle \rangle - 2
\spp \bigl\langle 
 w^{\alpha_1\beta_1}_{n_1n6}\bar{w}^{\beta_1\alpha_2}_{n_6n_3} 
w^{\alpha_2\beta_2}_{n_3n_8}\bar{w}^{\beta_2\alpha_1}_{n_8n_1} 
\bigr \rangle \Bigr ] .
\end{gather}
By using Wick theorem and Eqs.~\eqref{eq:prop:PH} - \eqref{eq:prop:PPT}, we find
\begin{gather}
[P_2^{\alpha_1\alpha_2}]^{(2)}(i\varepsilon_{n_1},i\varepsilon_{n_3})  = 
\left( \frac{64}{g}\right )^2\frac{\pi T}{g}\sum_{j=0}^3 \Gamma_j\int_{q,p} \notag \\ \times 
 \sum_{\omega_n>\varepsilon_{n_3}} \Bigr [ \mathcal{D}_q(i\omega_n+i\Omega^\varepsilon_{13}) +
 \mathcal{C}_q(i\omega_n+i\Omega^\varepsilon_{13}) \Bigl ] \notag \\ \times
\mathcal{D}_p(i\omega_n)\mathcal{D}^{(j)}_p(i\omega_n)
+ (\varepsilon_{n_1} \leftrightarrow \varepsilon_{n_3})
.
\label{eq2loopK2_P2++1}
\end{gather} 
Performing analytical continuation to real frequencies, $i\varepsilon_{n_1} \to E+i0^+$, $i\varepsilon_{n_3} \to E^\prime+i0^+$, we obtain
\begin{gather}
[P_2^{\alpha_1\alpha_2}]^{RR (2)}(E,E^\prime)  = 
 \left( \frac{32}{g}\right )^2\frac{1}{ig} \sum_{j=0}^3 \Gamma_j \int_{q,p,\omega} \mathcal{F}_{\omega-E^\prime}\notag \\
 \times \Bigl [ 
\mathcal{D}^R_q(\omega+E-E^\prime)+\mathcal{C}^R_q(\omega+E-E^\prime) \Bigr ] \notag \\ \times \mathcal{D}^R_p(\omega)
\mathcal{D}^{(j) R}_p(\omega)+ (E \leftrightarrow E^\prime).
\label{eq2loopK2_P2++11}
\end{gather} 
Setting $E=E^\prime=T=0$, we find (see Appendix \ref{app:sec:1})
\begin{equation}
[P_2^{\alpha_1\alpha_2}]^{RR (2)} \to 16 \frac{t^2\,h^{2\epsilon}}{\epsilon^2} \sum_{j=0}^3 \Bigl [
\ln (1+\gamma_j) -\frac{\epsilon}{4} \ln^2(1+\gamma_j)\Bigr ] ,
\label{eq2loopK2_P2++2}
\end{equation}
where we omit finite in $\epsilon$ terms. 

The two-loop contribution to $P^{\alpha_1\alpha_2}_2(i\varepsilon_{n_1},i\varepsilon_{n_4})$ can be written 
as follows
\begin{gather}
[P^{\alpha_1\alpha_2}_2]^{(2)}(i\varepsilon_{n_1},i\varepsilon_{n_2}) = 
-\frac{1}{4} \sum_{n_5n_6}\sum_{\beta_1\beta_2} \langle \langle \spp \bigl [
w^{\alpha_1\beta_1}_{n_1n_6}\bar{w}^{\beta_1\alpha_1}_{n_6n_1} \bigr ] \notag \\
\cdot
\spp \bigl [\bar{w}^{\alpha_2\beta_2}_{n_2n_5}{w}^{\beta_2\alpha_2}_{n_5n_2}  \bigr ]
\rangle \rangle - 2 \Bigl \langle \spp \bigl [w^{\alpha_1\alpha_2}_{n_1n_2} \bar{w}^{\alpha_2\alpha_1}_{n_2n_1}\bigr ] \notag \\
\times 
\Bigl [ S^{(4)}_\sigma+S^{(4)}_{\rm int}+\frac{1}{2} \left (S^{(3)}_{\rm int}\right )^2\Bigr ] \Bigr \rangle .
\label{eqP2+-0}
\end{gather} 
Here the term
\begin{align}
S^{(4)}_\sigma   = & -\frac{g}{128} \int_{q_j} \delta\left (\sum_{j=0}^3\bm{q_j}\right ) 
\sum_{\beta_1\beta_2\beta_3\beta_4}\sum_{n_5n_6n_7n_8}  \notag \\
& \times \spp \Bigl [ w^{\beta_1\beta_2}_{n_5n_6}(\bm{q_0}) \bar{w}^{\beta_2\beta_3}_{n_6n_7}(\bm{q_1})
w^{\beta_3\beta_4}_{n_7n_8}(\bm{q_2}) \bar{w}^{\beta_4\beta_1}_{n_8n_5}(\bm{q_3})\Bigl ]\notag \\
& \times
 \Bigl [ 2h^2+ \frac{16 z}{g} (\Omega^\varepsilon_{56}+\Omega^\varepsilon_{78})-(\bm{q_0}+\bm{q_1})(\bm{q_2}+\bm{q_3})
 \notag \\
 & -(\bm{q_0}+\bm{q_3})(\bm{q_1}+\bm{q_2}) \Bigr ] , 
\end{align}
appears in the expansion of $S_\sigma$ and the regulator term \eqref{SsGenFull} to the forth order in $W$. The expansion of the interaction term $S_{\rm int}$ results in the following third order term,
\begin{align}
S^{(3)}_{\rm int}  & = \frac{\pi T}{4} \sum_{r=0,3}\sum_{j=0}^3 \Gamma_j \sum_{\alpha,n} \int d\bm{r} \Tr I^{\alpha}_{n} t_{rj} W \notag \\
& \hspace{2.5cm} \times
\Tr I^{\alpha}_{-n} t_{rj}\Lambda W^2,
\end{align}
and forth order term,
\begin{align}
S^{(4)}_{\rm int} & = -\frac{\pi T}{16} \sum_{r=0,3}\sum_{j=0}^3 \Gamma_j \sum_{\alpha,n} \int d\bm{r} \Tr I^{\alpha}_{n} t_{rj} \Lambda W^2 \notag \\
& \hspace{2.5cm} \times 
\Tr I^{\alpha}_{-n} t_{rj}\Lambda W^2 .
\end{align} 
After evaluation of averages in Eq.~\eqref{eqP2+-0}, we find
\begin{widetext}
\begin{align}
& [P_2^{\alpha_1\alpha_2}]^{(2)}(i\varepsilon_{n_1},i\varepsilon_{n_2})  =  - \left ( \frac{16}{g}\right )^2 
\left [ \left ( \int_q \mathcal{D}_q(i\Omega^\varepsilon_{12})\right )^2 + \left ( \int_q \mathcal{C}_q(i\Omega^\varepsilon_{12})\right )^2 \right ] 
+ \frac{1-9}{4} \left ( \frac{16}{g}\right )^2 \int_{q,p} \Bigl [ p^2+q^2+h^2+\frac{16 z}{g} \Omega^\varepsilon_{12} \Bigr ]
\notag \\
& \hspace{0.5cm}
\times 
\mathcal{C}_p(i\Omega^\varepsilon_{12}) \mathcal{D}_q(i\Omega^\varepsilon_{12})\Bigl [\mathcal{D}_q(i\Omega^\varepsilon_{12})+\mathcal{C}_p(i\Omega^\varepsilon_{12}) \Bigr ] 
 - \left ( \frac{64}{g}\right )^2 \sum_{j=0}^3 \frac{\pi T\Gamma_j}{g} \int_{q,p} \Bigl [ \mathcal{D}^2_p(i\Omega^\varepsilon_{12})+\mathcal{C}^2_p(i\Omega^\varepsilon_{12})\Bigr ] 
\Bigl \{ \sum_{\omega_n>\varepsilon_{n_1}} + \sum_{\omega_n>-\varepsilon_{n_2}}\Bigr \}
\notag \\
& \hspace{0.5cm} \times 
 \Bigl [ p^2+q^2 +2h^2 +\frac{16 z}{g} \bigl ( \Omega^\varepsilon_{12} + \omega_n \bigr ) \Bigr ] \mathcal{D}_q(i\omega_n) \mathcal{D}^{(j)}_q(i\omega_n)
   + \left ( \frac{64}{g}\right )^2 \sum_{j=0}^{3} \frac{2\pi T\Gamma_j}{g} \int_{q,p} \sum_{\omega_n>0} 
   \Bigl [1 - \frac{16 \Gamma_j \omega_n}{g} \mathcal{D}_{\bm{q}+\bm{p}}^{(j)}(i\omega_{n})  \Bigr ]
   \notag \\
& \hspace{0.5cm}   \times
\Bigl [ \mathcal{D}^2_q(i\Omega^\varepsilon_{12}) \mathcal{D}_p(i\Omega^\varepsilon_{12}+i\omega_n) + \mathcal{C}^2_q(i\Omega^\varepsilon_{12}) \mathcal{C}_p(i\Omega^\varepsilon_{12}+i\omega_n) \Bigr ]
 + \left ( \frac{64}{g}\right )^2 \sum_{j=0}^{3} \frac{\pi T\Gamma_j}{g} \int_{q,p} \Bigl \{\sum_{\varepsilon_{n_1}>\omega_n>0} 
 + \sum_{-\varepsilon_{n_2}>\omega_n>0} \Bigr \}
 \notag \\
& \hspace{0.5cm}   \times
   \Bigl [1 - \frac{16 \Gamma_j \omega_n}{g} \mathcal{D}_{\bm{q}+\bm{p}}^{(j)}(i\omega_{n})  \Bigr ]
   \Bigl [ \mathcal{D}^2_q(i\Omega^\varepsilon_{12}) \mathcal{D}_p(i\Omega^\varepsilon_{12}-i\omega_n) + \mathcal{C}^2_q(i\Omega^\varepsilon_{12}) \mathcal{C}_p(i\Omega^\varepsilon_{12}-i\omega_n) \Bigr ]
 .
\label{eq2loopK2_P2+-1}
\end{align}
Performing analytic continuation to the real frequencies, $i\varepsilon_{n_1} \to E+i0^+$, $i\varepsilon_{n_2} \to E^\prime-i0^+$, in Eq. \eqref{eq2loopK2_P2+-1}, we obtain
\begin{align}
& [P_2^{\alpha_1\alpha_2}]^{RA (2)}(E,E^\prime) =  
- \left ( \frac{16}{g}\right )^2 
\left [ \left ( \int_q \mathcal{D}^R_q(\Omega)\right )^2 + \left ( \int_q \mathcal{C}^R_q(\Omega)\right )^2 \right ] 
-2 \left (\frac{16}{g}\right )^2 \int_{q,p} \Bigl [ p^2+q^2+h^2-\frac{16 z}{g} i \Omega \Bigr ]
\mathcal{C}^R_p(\Omega) 
\notag \\
& \hspace{0.5cm}
\times 
\mathcal{D}^R_q(\Omega) \Bigl [\mathcal{D}^R_q(\Omega)+\mathcal{C}^R_p(\Omega) \Bigr ] 
 - \left ( \frac{32}{g}\right )^2 \sum_{j=0}^3 \frac{\Gamma_j}{i g} \int_{q,p,\omega} \Bigl [ \mathcal{D}^{R2}_p(\Omega)+\mathcal{C}^{R2}_p(\Omega)\Bigr ] 
\Bigl [ \mathcal{F}_{\omega-E} + \mathcal{F}_{\omega+E^\prime}\Bigr ]
\mathcal{D}^R_q(\omega) \mathcal{D}^{(j)R}_q(\omega)
\notag \\
& \hspace{0.5cm} \times 
 \Bigl [ p^2+q^2 +2h^2 - \frac{16 z}{g} i \bigl ( \Omega + \omega \bigr ) \Bigr ] 
 + \left ( \frac{64}{g}\right )^2 \sum_{j=0}^{3} \frac{\Gamma_j}{2i g} \int_{q,p,\omega}  \mathcal{B}_\omega
  \Bigl [ \mathcal{D}^{R2}_q(\Omega) \mathcal{D}^R_p(\omega+\Omega) + \mathcal{C}^{R2}_q(\Omega) \mathcal{C}^R_p(\omega+\Omega) \Bigr ]
\notag \\
&  \hspace{0.5cm} \times \Bigl [1 + \frac{16 \Gamma_j i \omega}{g} \mathcal{D}_{\bm{q}+\bm{p}}^{(j) R}(\omega)  \Bigr ]
+ \left ( \frac{32}{g}\right )^2 \sum_{j=0}^{3} \frac{\Gamma_j}{i g} \int_{q,p,\omega} \Bigl [ 2 \mathcal{B}_\omega - \mathcal{F}_{\omega-E} - \mathcal{F}_{\omega+E^\prime}\Bigr ] \Bigl [1 + \frac{16 \Gamma_j i \omega}{g} \mathcal{D}_{\bm{q}+\bm{p}}^{(j) R}(\omega)  \Bigr ]
\notag \\
&  \hspace{0.5cm} \times
  \Bigl [ \mathcal{D}^{R2}_q(\Omega) \mathcal{D}^R_p(\Omega-\omega) + \mathcal{C}^{R2}_q(\Omega) \mathcal{C}^R_p(\Omega-\omega) \Bigr ]
.
\label{eq2loopK2_P2+-1}
\end{align}
\end{widetext}
Here we introduce the bosonic distribution function  $\mathcal{B}_\omega = \coth (\omega/2T)$. 
We note that the most part of the two-loop contribution to $[P_2^{\alpha_1\alpha_2}]^{RA}(E,E^\prime)$ can be considered as the renormalization of the diffuson and cooperon which determine one-loop contribution to  $[P_2^{\alpha_1\alpha_2}]^{RA}(E,E^\prime)$ (see Appendix \ref{app:sec:2}).
Again setting $E=E^\prime=T=0$,  we derive (see Appendix \ref{app:sec:1})
\begin{gather}
[P_2^{\alpha_1\alpha_2}]^{RA (2)} \to - 32 \frac{t^2\,h^{2\epsilon}}{\epsilon^2} \Bigl [ 3+\epsilon \Bigr ]- 16 \frac{t^2\,h^{2\epsilon}}{\epsilon^2}\sum_{j=0}^3\Bigl [
2 f(\gamma_j ) \notag \\
+ 3 \ln (1+\gamma_j) 
 -
\epsilon\frac{2+\gamma_j}{\gamma_j}  \Bigl ( \ln(1+\gamma_j) + \liq(-\gamma_j) \notag \\
+ \frac{1}{4}\ln^2(1+\gamma_j)\Bigr ) \Bigr ] ,
\label{eq2loopK2_P2+-2}
\end{gather}
where 
\begin{equation}
\label{fx}
f(x) = 1 - (1+1/x)\ln(1+x)
\end{equation}
 and $\liq(x) = \sum_{k=1}^\infty x^k/k^2$ denotes the polylogarithm.
Combining together Eqs. \eqref{eq2loopK2_P2++2} and \eqref{eq2loopK2_P2+-2}, we obtain the following two-loop  
contribution to the irreducible two-point correlation function:
\begin{align}
K_2^{(2)}  & = \rho_0^2 \frac{t^2\,h^{2\epsilon}}{\epsilon^2} \Biggl \{ 1+2 \Bigl (1 +\frac{\epsilon}{2}\Bigr ) + \sum_{j=0}^3 \Bigl [   f(\gamma_j) + 2 \ln(1+\gamma_j) \notag \\
&  + \frac{\epsilon}{2} \bigl [ \ln(1+\gamma_j) + 2 f(\gamma_j) - c(\gamma_j) 
\bigr ]\Bigr ] \Biggr \} ,\label{eqK1_0}
\end{align}
where we introduced the function
\begin{gather}
c(\gamma) = 2 +\frac{2+\gamma}{\gamma} \liq(-\gamma) + \frac{1+\gamma}{2\gamma} \ln^2(1+\gamma) .
\label{eq:def:c}
\end{gather}

\subsubsection{Anomalous dimension}

It is well-known (see e.g., Ref. [\onlinecite{baranov99}]) that the momentum scale $h$ acquires renormalization. The corresponding renormalized momentum scale  $h^\prime$ can be defined as follows
\begin{equation}
g^\prime h^{\prime 2}  \Tr \Lambda^2 = g h^2 \langle \Tr \Lambda Q \rangle , 
\end{equation}  
where $g^\prime$ denotes renormalized conductivity at the momentum scale $h^\prime$. In the one-loop approximation, one can find [\onlinecite{baranov99}] 
\begin{equation}
h^\prime = h \Biggl \{  1 - \frac{t\,  h^\epsilon}{2\epsilon} \Bigl [ 1+ \sum_{j=0}^3\bigl [ f(\gamma_j)+\frac{1}{2}\ln(1+\gamma_j) \bigr ] \Bigr ] \Biggr \} 
\label{eqhren}
\end{equation}
and [\onlinecite{AA1979AAL1980,Fin198384,Castellani1984}] 
\begin{gather}
g^\prime = g \Bigl [1+\frac{a_1 t\, h^\epsilon}{\epsilon} +O(\epsilon)\Bigr ], \quad
 a_1 =1+\sum_{j=0}^3 f(\gamma_j). 
\label{eqS1}
\end{gather}
We mention that $g^\prime h^{\prime 2} = g h^2 Z^{1/2}$ as expected.  

By using Eq. \eqref{eqhren}, we can write the second moment of the local density of states in terms of the renormalized momentum scale $h^\prime$  and factor $Z$ as follows:
\begin{equation}
\langle \rho^2 \rangle = Z \rho_0^2 +K_2 = \rho_0^2 Z m^\prime_2, 
\end{equation}
where, keeping only terms with pole structure in $\epsilon$,
\begin{equation}
m^\prime_2 =  m_2 \Bigl [ 1+ \frac{b^{(2)}_1 t \, h^{\prime \epsilon}}{\epsilon}+ \frac{t^2 h^{\prime 2\epsilon}}{\epsilon^2} \Bigl (b^{(2)}_2+\epsilon b^{(2)}_3\Bigr ) \Bigr ] .
\label{eqM2def}
\end{equation}
Here $m_2=1$ and 
\begin{gather}
b^{(2)}_1 = -2, \quad b^{(2)}_2 = 3+\sum_{j=0}^3 f(\gamma_j), 
\quad b^{(2)}_3 = - \frac{1}{2} \sum_{j=0}^3 c(\gamma_j) .
\end{gather} 

In order to find the anomalous dimension of $m^\prime_2$, we introduce dimensionless quantity $\bar{t} = t^\prime h^{\prime \epsilon}$ and, using Eqs \eqref{eqS1} and \eqref{eqM2def}, express $t$, $\gamma_j$ and $m_2$  as
\begin{gather}
t =  (h^\prime)^{-\epsilon} \bar{t} Z_t(\bar{t},\gamma^\prime_s,\gamma^\prime_t),\qquad \gamma_j = \gamma_j^\prime Z_j(\bar{t},\gamma^\prime_s,\gamma_t^\prime), \notag \\
 m_2= m_2^\prime 
Z_{m_2}(\bar{t},\gamma^\prime_s,\gamma^\prime_t) .
\end{gather}
We remind that interaction parameters $\gamma_j$ are renormalized at the one-loop level. However,
since $b^{(2)}_1$ is independent of $\gamma_j$ this renormalization does not affect the two-loop result 
for the anomalous dimension  of $m^\prime_2$. To the lowest orders in $\bar{t}$ the renormalization parameters become
\begin{equation}
Z_t= 1 + \frac{a_1}{\epsilon}\bar{t}, 
\label{eqZZZ2}
\end{equation}
and
\begin{equation}
Z_{m_2}^{-1} =  1 + \frac{b^{(2)}_1}{\epsilon}\bar{t} +
\frac{\bar{t}^2}{\epsilon^2} \Bigl [ b^{(2)}_2 + b^{(2)}_1 a_1 + \epsilon b^{(2)}_3 \Bigr ] .
\label{eqZZZ2}
\end{equation}
Now the renormalization group function for $m_2^\prime$ can be derived in a standard manner from the conditions that $m_2$ (as well as $t$ and $\gamma_j$) does not depend on the momentum scale $h^\prime$. Thus, we obtain the two-loop result for the anomalous dimension $\zeta_2(t,\gamma_s.\gamma_t)$ of $m_2$:
\begin{equation}
-\frac{d \ln m_2}{d \ln y} = \zeta_2 =  -2 t - [c(\gamma_s)+3c(\gamma_t)] t^2 + O(t^3) .
\label{eqm2RG}
\end{equation}
Here $y=1/h^\prime$ is the renormalization group running length scale and we omit `prime' and `bar' signs for a brevity. The function $c(\gamma)$ is defined in Eq. \eqref{eq:def:c}.
It is worthwhile to mention that $c(0) = 0$ as it is known for free electrons [\onlinecite{Wegner79}], and $c(-1) = 2 -\pi^2/6\approx 0.36$. Remarkably, the interaction affects the anomalous dimension at the two-loop order only.  
We emphasize that the relation $b^{(2)}_2 = b^{(2)}_1(b^{(2)}_1-a_1)/2$ guaranties the renormalizability of $m_2$, i.e. the absence in Eq.~\eqref{eqm2RG} of terms divergent in the limit $\epsilon\to 0$.  {In addition, this indicates also that the operator corresponding to $K_2$ is the eigenoperator under action of the renormalization group. Indeed, if the operator corresponding to $K_2$ consists of several eigenoperators, the relation $b^{(2)}_2 = b^{(2)}_1(b^{(2)}_1-a_1)/2$ would imply non-linear system of equations which has no non-trivial solutions in general.}

\begin{figure}[t]
\centerline{\includegraphics[width=80mm]{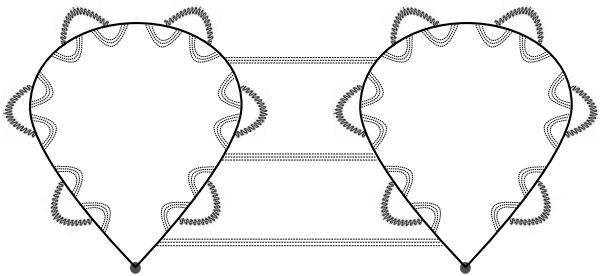}}
\caption{Representative diagram for the two-point correlation function of the local density of states. Solid lines denote electron Green functions, while wavy lines denote the dynamically screened Coulomb interaction. Ladders of dashed lines represent diffusons and cooperons. }
\label{Figure1}
\end{figure}

To illustrate the two-loop contribution to the renormalization group equation \eqref{eqm2RG} we show in Fig. \ref{Figure1} representative diagrams for the two-point correlation function of the local density of states. Each local density of states is given by a fermionic loop dressed by interaction lines. Averaging each loop over disorder generates diffusive vertex corrections and yields the suppression of the average local density of states by gauge-type phase fluctuations. On the other hand, diffusons and cooperons connecting the loops lead to multifractal correlations. 

The results of this section imply that the second moment of the local density of states at $E=T=0$ can be written as 
\begin{equation}
\langle \rho^2 \rangle = \langle \rho \rangle^2\, m^\prime_2  ,
\label{eqK3}
\end{equation}
where the scaling behavior of $m_2$ is governed by Eq.~\eqref{eqm2RG}. We mention that the interaction affects the anomalous dimension $\zeta_2$ only at the two-loop level.

\subsection{The $q$-th moment of the local density of states}
 
In this section we demonstrate that in the two-loop approximation the $q$-th moment of the local density of states at $E=T=0$ can be written as 
\begin{equation}
\left \langle \rho^q\right \rangle =\langle \rho\rangle^q m^\prime_q  ,
\label{eqKmq}
\end{equation}
where the behavior of $m_q$ for the orthogonal case (both time reversal and spin rotational symmetries are preserved) is determined by the following renormalization group equation:
\begin{equation}
-\frac{d \ln m_q}{d \ln y} = \zeta_q = \frac{q(1-q)}{2} \Bigl \{ 2 t +  \bigl [c(\gamma_s) +3c(\gamma_t) \bigr ] t^2 \Bigr \}+ O(t^3) .
\label{eqmqRG}
\end{equation}
Here the function $c(\gamma_j)$ is given in Eq. \eqref{eq:def:c}. We mention that Eq. \eqref{eqmqRG} implies 
\begin{equation}
m^\prime_q =  m_q \Bigl [ 1+ \frac{b_1^{(q)} t \, h^{\prime \epsilon}}{\epsilon}+ \frac{t^2 h^{\prime 2\epsilon}}{\epsilon^2} \bigl (b^{(q)}_2+\epsilon b^{(q)}_3\bigr )\Bigr ] 
\label{eqmqe}
\end{equation}
with $m_q=1$ and
\begin{gather}
b_1^{(q)}= \frac{q(q-1)}{2} b_1^{(2)}, \qquad b_2^{(q)}= \frac{b_1^{(q)}}{2} (b_1^{(q)} - a_1) , \notag \\
b_3^{(q)}= \frac{q(q-1)}{2} b_3^{(2)}  .
\label{eqmqe1}
\end{gather}
Let us consider the  irreducible $q$-th moment of the local density of states, $K_q = \bigl \langle \bigl ( \rho - \langle \rho \rangle \bigr )^q \bigr \rangle$ (with $q\geqslant 3$). Then we can write
\begin{equation}
\left \langle \rho^q \right \rangle  = \sum\limits_{j=0}^{q-1} (-1)^{q-1-j} C^q_j  \langle \rho^j\rangle \langle \rho \rangle^{q-j} + K_q ,
\end{equation}
where $C_j^q = q!/[(q-j)!j!]$. Provided Eqs \eqref{eqKmq} and \eqref{eqmqe} hold for all $0\leqslant j\leqslant q-1$, we find 
\begin{equation}
\left \langle \rho^q \right \rangle =
Z^{q/2} \tilde{m}_q + K_q, 
\end{equation}
where
\begin{equation} 
\tilde{m}_q =  m_q\Bigl [ 1+ \frac{\tilde{b}_1^{(q)} t \, h^{\prime \epsilon}}{\epsilon}+ \frac{t^2 h^{\prime 2\epsilon}}{\epsilon^2} \bigl (\tilde{b}^{(q)}_2+\epsilon \tilde{b}^{(q)}_3 \bigr )\Bigr ] ,
\label{eqmqe3}
\end{equation}
with
\begin{equation}
\tilde{b}_1^{(q)}= \frac{k_q}{2} b_1^{(2)}, \,  \tilde{b}_2^{(q)}= 
\frac{b_1^{(2)}[l_q b_1^{(2)} - 2 k_q a_1]}{8}
, \, \tilde{b}_3^{(q)}= \frac{k_q}{2} b_3^{(2)} .
\end{equation}
The combinatorial coefficients $k_q$ and $l_q$ are defined via derivatives of the function $\mathcal{P}_q(x)=x^q-(x-1)^q$ at the point $x=1$:
\begin{equation}
k_q  = \mathcal{P}_q^{\prime\prime}(1) , \quad
l_q  = \Bigl ( x^2 \mathcal{P}_q^{\prime\prime}(x)\Bigr )^{\prime\prime}\Bigl 
|_{x=1}  .
\label{eqmqe4}
\end{equation}
As one can check, $k_q = q(q-1)$ for $q\geqslant 3$ whereas $l_q=k_q^2$ for $q\geqslant 5$. For $q=3$ and $q=4$ one finds $l_3=12$ and $l_4=120$. Since expression for $K_q$ involves connected contributions from averages of the number $q$ of matrices $Q$, there is no two-loop contribution to $K_q$ for $q\geqslant 5$. Therefore, with the help of Eq. \eqref{eqmqe4}, we obtain the result \eqref{eqmqe} for $q\geqslant 5$. The cases $q=3$ and $q=4$ are needed to be considered specially. One can demonstrate that Eq. \eqref{eqmqRG}
 holds for $q=3$ and $q=4$ also (see Appendix \ref{app:sec:3}).

\section{Scaling analysis}
\label{s4}
 
\subsection{General scaling results} 
 \label{s4.1}
 
Near the interacting critical point Eqs.~\eqref{eqKmq} and \eqref{eqmqRG} imply that at zero energy and temperature, $E=T=0$, the $q$-th moment of the local density of states obeys the following scaling behavior
\begin{equation}
\langle \rho^q \rangle \sim \langle \rho \rangle^q \left ({\xi}/{l}\right )^{-\Delta_q} \Upsilon_q(\xi/L) .
\label{Scal_eq3}
\end{equation}
Here $L$ stands for the system size, $l$ and $\xi= l |1-t/t_*|^{-\nu}$ denote the mean free path and the correlation length, respectively. The multifractal critical exponent $\Delta_q$ is determined by the anomalous dimension of $m_q$ at the critical point, $\Delta_q = \zeta_q^*$ ($q\geqslant 2$). We note that $\Delta_1 =0$ by definition.
The scaling function $\Upsilon_q(x)$ has the following asymptotes
\begin{equation}
\Upsilon_q(x) = \begin{cases}
1, & \qquad x\ll 1 ,\\
x^{\Delta_q}, & \qquad x\gg 1 .
\end{cases} 
\label{Scal_eq10}
\end{equation}

As one can see from Eq. \eqref{Scal_eq3}, the scaling behavior of the $q$-th moment of the local density of states is determined also by the scaling behavior of the average DOS. At zero energy and temperature, $E=T=0$, one can write  from Eq. \eqref{eq:ad:Z} (see e.g., [\onlinecite{finkelstein90,belitz94}]) 
\begin{equation}
\langle \rho \rangle \sim (\xi/l)^{-\theta} {\Upsilon}(\xi/L), 
\label{Scal_eq2}
\end{equation}
where the critical exponent $\theta$ is determined by the anomalous dimension of the disorder-averaged local density of states at the critical point, $\theta=\zeta^*$. The scaling function ${\Upsilon}$ behaves as follows
 \begin{equation}
{\Upsilon}(x) = \begin{cases}
1, & \qquad x\ll 1 , \\
x^{\theta}, & \qquad x\gg 1 .
\end{cases} 
\end{equation}
Combining Eqs \eqref{Scal_eq3} and \eqref{Scal_eq2}, we find
\begin{equation}
\langle \rho^q \rangle \sim  \left ({\xi}/{l}\right )^{-\theta q-\Delta_q} \tilde{\Upsilon}_q(\xi/L) ,
\label{Scal_eq4}
\end{equation}
where the scaling function $\tilde\Upsilon_q(x)$ has the following properties:
\begin{equation}
\tilde\Upsilon_q(x) = \begin{cases}
1, & \qquad x\ll 1 , \\
x^{\Delta_q+\theta q}, & \qquad x\gg 1 .
\end{cases} 
\end{equation}
In general, the exponent, $\theta$ is positive, which corresponds to a suppression of the average tunneling density of states. While our calculation of moments is restricted to integer positive $q$, the results for the multifractal exponents $\Delta_q$ can be extended (by analytic continuation) to all real (and, in fact, even complex) $q$. Indeed, the local density of states $\rho(\bf r)$ is a real positive quantity and thus the moments $\langle\rho^q\rangle$ are unambiguously defined for any $q$. By definition $\Delta_0=\Delta_1=0$. According to general properties of the multifractality spectra, $d^2\Delta_q/dq^2<0$, so that $\Delta_q$ is positive for $0<q<1$ and negative on the rest of the real axis. The combination $\theta q+\Delta_q$ controlling the scaling of moments $\langle\rho^q\rangle$ (without normalization to the average) is positive for not too large positive $q$. It is expected that the absolute value of $\Delta_q$ grows sufficiently fast, so that $\theta q+\Delta_q$ becomes negative at $q>q_c$ with some $q_c>1$. This means that although the average local density of states is suppressed, its sufficiently high moments are enhanced (in comparison with a clean system) by a combined effect of interaction and disorder. 

Interestingly, the results \eqref{Scal_eq3}, \eqref{Scal_eq4} are similar to the behavior of the local density of states at critical points in non-interacting systems of unconventional symmetry classes (see
Ref. [\onlinecite{Evers08}] and references therein). However, the physics in the two
cases is essentially different. In unconventional symmetry classes, suppression of the local density of states occurs near a special point of the single-particle spectrum due to existence of an additional symmetry. For example, it is the case for non-interacting Dirac fermions subjected to special types of disorder.
Contrary to this, in our problem the suppression of disorder-averaged
local density of states takes place because of Coulomb interaction and, therefore, is pinned to the chemical
potential. This suppression is a genuine many-body effect that has
common roots with formation of gap in Mott insulators and in Coulomb-blockade
regime of quantum dots, as well as of soft Coulomb gap [\onlinecite{ES75SE84}] in disordered
insulators.

As usual, in the presence of interactions, finite energy or temperature induces the inelastic length $L_\phi$ related with the  dephasing time $\tau_\phi$: $L_\phi \sim \tau_\phi^{1/z}$, where $z$ is the dynamical exponent. In the case of Coulomb interaction the frequency/energy and temperature scaling are the same such that $1/\tau_\phi \sim \max\{|E|,T\}$. Therefore, the inelastic length  becomes $L_\phi \sim \min\{L_E, L_T\}$ with $L_E\sim |E|^{-1/z}$ and $L_T\sim T^{-1/z}$. We emphasize that the energy $E$ is counted from the chemical potential. Provided $L_\phi \ll L$, the inelastic length should be substituted for $L$ in Eqs.~\eqref{Scal_eq3}, \eqref{Scal_eq2} and \eqref{Scal_eq4}. Therefore, our scaling results for the $q$-th moment can be summarized as follows:
\begin{equation}
\begin{split}
\langle \rho^q(E,\bm{r}) \rangle & \sim \langle \rho(E) \rangle^q \bigl ({\mathcal{L}}/{l}\bigr )^{-\Delta_q} , \\
 \langle \rho(E) \rangle & \sim \bigl ({\mathcal{L}}/{l}\bigr )^{-\theta} ,
 \end{split}
\label{Scal_eq5}
\end{equation}
where $\mathcal{L} = \min\{L,\xi,L_\phi\}$. Note that the exponent $\theta$ is related with the exponent $\beta$ which determines the energy dependence of the disorder-average local density of states at the criticality, $\langle \rho(E)\rangle \sim |E|^{\beta}$, as $\beta= \theta/z$. 

The scaling behavior of the 2-point correlation function of the local density of states at the same energy but different spatial points is controlled by the exponent $\Delta_2$ also. At $l< R < \mathcal{L}$, we find the following scaling:
\begin{equation}
\langle \rho(E,\bm{r}) \rho(E,\bm{r}+\bm{R})\rangle \sim \langle \rho(E) \rangle^2 \bigl (R/\mathcal{L} \bigr )^{-\eta} .
\label{Scal_eq6}
\end{equation}
Here the exponent $\eta=-\Delta_2$. This result follows from two observations: (i) at $R\sim l$ Eq. \eqref{Scal_eq6} should reproduce Eq.~\eqref{Scal_eq5} with $q=2$, (ii) at $R \gtrsim \mathcal{L}$ the local density of states at points $\bm{r}$ and $\bm{r}+\bm{R}$ is essentially uncorrelated. Similarly (see, e.g. Ref. [\onlinecite{Evers08}]), one can find at $l\ll R \ll \mathcal{L}$ that  
\begin{gather}
\langle \rho^{q_1}(E,\bm{r}) \rho^{q_2}(E,\bm{r}+\bm{R})\rangle  \sim  \langle \rho(E) \rangle^{q_1+q_2} 
\bigl (\mathcal{L}/l\bigr )^{-\Delta_{q_1}-\Delta_{q_2}} 
\notag \\
\times \bigl (R/\mathcal{L} \bigr )^{\Delta_{q_1+q_2}-\Delta_{q_1}-\Delta_{q_2}}  .
\end{gather}

The 2-point correlation function of the local density of states at the same spatial point but at different energies shows the following scaling behavior for $l < L_\omega, L_E < \min\{L, \xi, L_T\}$:
\begin{gather}
\frac{\langle \rho(E,\bm{r})  \rho(E+\omega,\bm{r}) \rangle}{\langle \rho(E) \rangle
\langle \rho(E+\omega) \rangle} \sim 
 \bigl (L_\omega/{l}\bigr )^{\eta}\hat{\Upsilon}_2(L_\omega/L_E) .
 \label{Scal_eq11}
 \end{gather}
Here $L_\omega \sim |\omega|^{-1/z}$ and the scaling function  $\hat{\Upsilon}_2(x)$ has the same asymptotes as the function  $\Upsilon_2(x)$    (see Eq. \eqref{Scal_eq10}).  In the case $l< L_\omega < \mathcal{L}$, one expects scaling as follows
\begin{equation}
\langle \rho(E,\bm{r})  \rho(E+\omega,\bm{r}) \rangle \sim \bigl ({\mathcal{L}}/l\bigr )^{-\theta}
 \bigl (L_\omega/{l}\bigr )^{-\theta+\eta} .
 \label{Scal_eq12}
\end{equation}

Next, using Eqs. \eqref{Scal_eq6} and \eqref{Scal_eq11}, we find the following scaling behavior of the 2-point correlation function of the local density of states at different energies and different spatial points in the most interesting case $l < R < L_\omega < \mathcal{L}$:
\begin{gather}
\frac{\langle \rho(E,\bm{r})  \rho(E+\omega,\bm{r}+\bm{R}) \rangle}{\langle \rho(E) \rangle
\langle \rho(E+\omega) \rangle} \sim 
 \bigl (L_\omega/{R}\bigr )^{\eta}.
\label{Scal_eq14}
\end{gather}
 
\begin{figure}[t]
\centerline{\includegraphics[width=40mm]{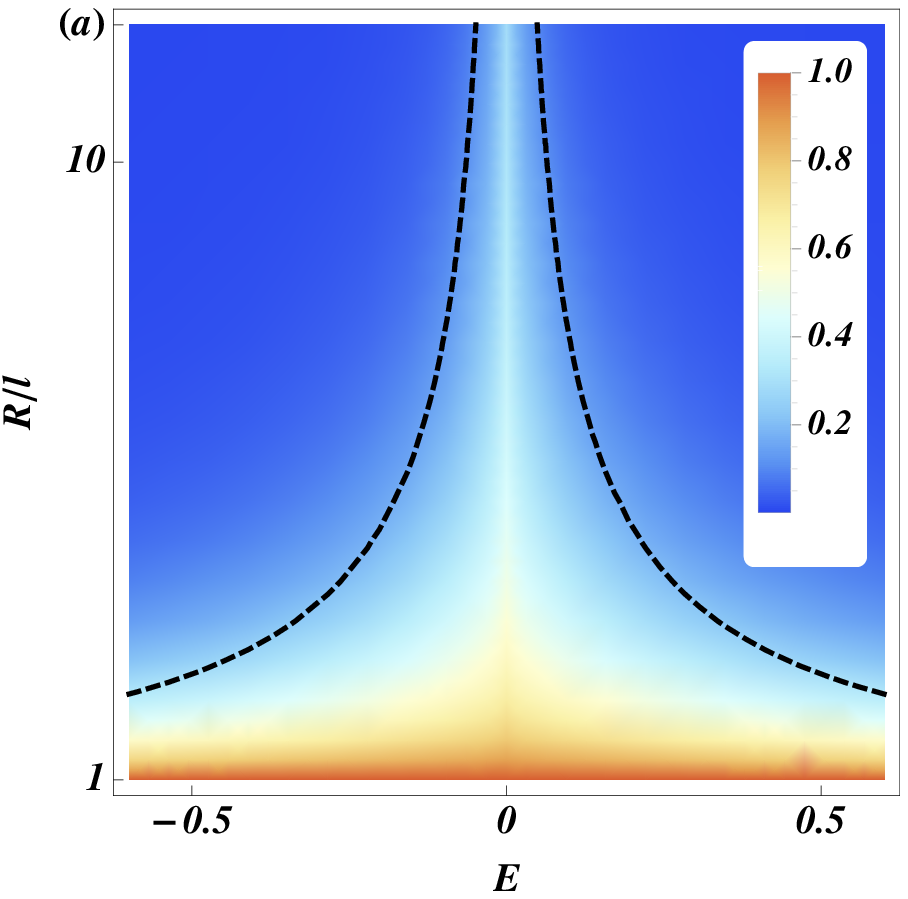}\hspace{.5cm}\includegraphics[width=40mm]{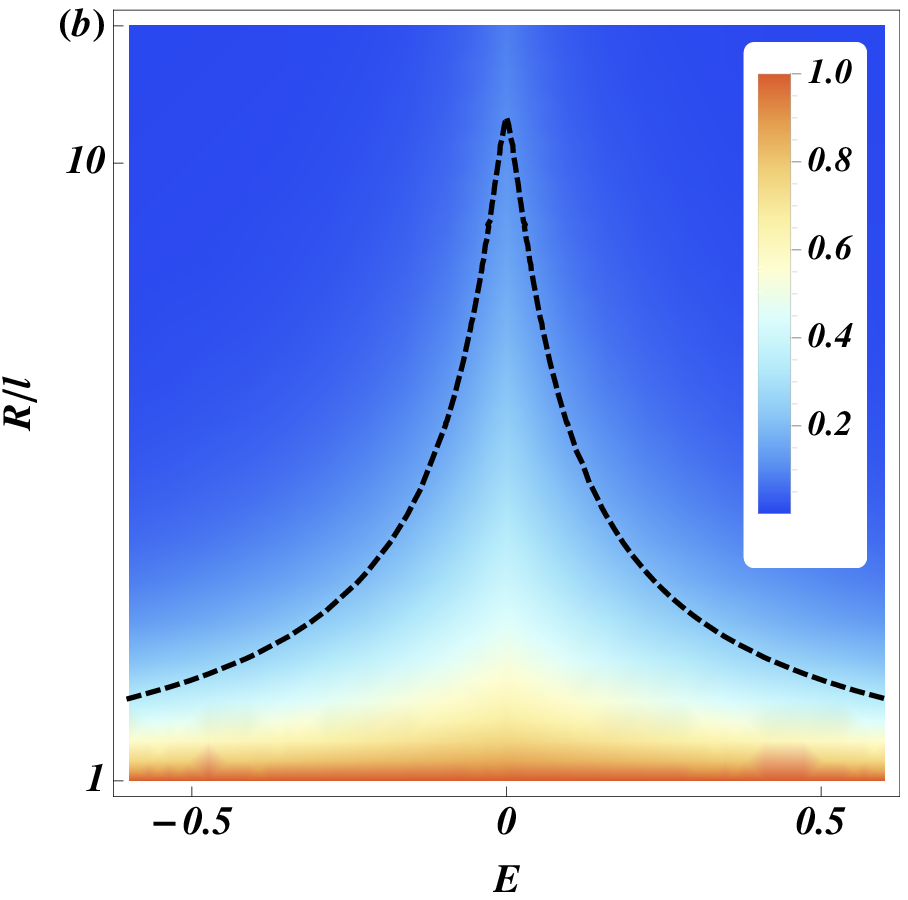}}
\caption{(Color online) Schematic color-code plot of the normalized autocorrelation
function $K_2(E,\bm{r},E,\bm{r}+\bm{R})/\langle\rho(E)\rangle^2$ for the system a) at the critical point,
$t=t_*$, and b) slightly on the metallic side, $(t_*-t)/t_*=0.15$. The energy is
measured in units of elastic scattering rate which sets the ultraviolet
cutoff of the NLSM theory. The dashed curve in a) and b) corresponds to
$R/l\sim(|E|^{2/z}+(1-t/t_*)^{2\nu})^{-1/2}$. }
\label{Figure2a2b}
\end{figure}

To visualize the spatial correlations \eqref{Scal_eq14} of the local density of states near a
metal-insulator transition, we present in Fig. \ref{Figure2a2b} a color-code plot of the normalized autocorrelation
function
$K_2(E,\bm{r},E,\bm{r}+\bm{R})/\langle\rho(E)\rangle^2$ (see Eq. \eqref{eq:K2:def:gen}). For this plot, we have chosen the following values of the critical exponents: $\nu=1$, $z=1.5$, $\eta=0.5$, which are
theoretical estimates obtained by taking $\epsilon=1$ in the one-loop results
for the case of Coulomb interaction with broken time reversal symmetry and spin invariance (see Sec. \ref{s4.2} below). The left panel (Fig. \ref{Figure2a2b}a) corresponds to the case when the system is exactly at the
transition point, $t=t_*$. We see the long-range multifractal correlation around the Fermi level, $E=0$. The values of critical exponents we use satisfy the inequality $\nu>1/z$. This inequality holds for experimental estimates of the corresponding exponents at 3D metal-insulator transitions and at quantum
Hall transitions. Then at zero temperature the range of correlations, which is given by $\mathcal{L}$, is controlled by the dephasing length $L_\phi \sim |E|^{-1/z}$. In the right panel (Fig. \ref{Figure2a2b}b), the system is slightly on the metallic side of the transition, $t<t_*$. In this case the range of correlations in the local density of states $\mathcal{L}$ is governed by the correlation length $\xi$ for a
certain window around the chemical potential, $|E| \lesssim \Delta$. On the metallic side of the transition the energy scale $\Delta \sim \xi^{-1/z} \sim (t_*-t)^{\nu z}$ determines the critical region near interacting critical point. Interestingly, that on insulator side of transition similar energy scale controls the position of the mobility edge for single-particle excitations [\onlinecite{BGM2014}]. Away from the Fermi level, $|E| \gtrsim \Delta$ the range of correlations is controlled by the dephasing length $L_\phi$. All essential
features of Fig. \ref{Figure2a2b} compare well with  Figs. 4A and 4B of the experimental paper~[\onlinecite{richardella10}]. Provided the inequality $\nu z>1$ is fulfilled, the spatial correlations of the local density of states look similarly to Fig. \ref{Figure2a2b} for any metal-insulator transition in the presence of Coulomb interaction. 

We are now going to evaluate the multifractal exponents in various symmetry classes. We limit ourselves to conventional classes (i.e., no particle-hole symmetries), which are classified by the presence or absence of time reversal and spin-rotational (full or partial) symmetries. We will first explore in Sec.~\ref{s4.2}, \ref{s4.4}, and \ref{s4.5} the classes with broken (at least partly) spin rotation invariance. In these classes $d=2$ serves as a lower critical dimension for the Anderson transition, so that the transition in  $d=2+\epsilon$ dimensions with small $\epsilon$ takes place in the weak-coupling regime  and can be controllably studied within the $\epsilon$-expansion.  We will also briefly discuss critical points relevant to disordered topological insulators for which $\sigma$-model action is supplemented by terms of topological character.  We will then turn in Sec.~\ref{s4.3} to systems with preserved spin-rotation invariance. In this case there is no weak-coupling Anderson transition in  $d=2+\epsilon$ dimensions since the triplet-channel interaction grows under renormalization. This shifts the transition in the range of intermediate couplings. The conclusion is valid also for 2D systems. We will estimate multifractal exponents for this 2D metal-insulator transition.

\subsection{Broken time reversal and spin rotational symmetries}
  \label{s4.2}
 
As the first example, we consider a system of disordered fermions with Coulomb interaction in the absence of time reversal and spin rotational symmetry, which corresponds to the symmetry class ``MI(LR)'' in terminology of
Ref.~\onlinecite{belitz94}. In the absence of electron-electron interactions this is the unitary Wigner-Dyson class A. 
For example, such situation occurs in the presence of magnetic impurities. In this case, all cooperon modes ($W$ with $r=1,2$) and triplet diffuson modes ($W$ with $j=1,2,3$ and $r=0,3$) are suppressed at large length scales.  The anomalous dimension of the $q$-th moment of the local density of states becomes [\onlinecite{BGM2013}]
\begin{equation}
-\frac{d \ln m_q}{d \ln y} = \zeta_q = \frac{q(1-q)}{2} \Bigl [ \frac{t}{2} +  c(-1) \frac{t^2}{4}  \Bigr ]+ O(t^3) .
\label{eqmqRG-U}
\end{equation}
Here the function $c(\gamma)$ is defined by Eq. \eqref{eq:def:c} such that $c(-1)= 2-\pi^2/6$.
We note the factor-of-$2$ difference in definition of $t$ in Ref. [\onlinecite{BGM2013}]. The two-loop renormalization group analysis of the Anderson transition in $d=2+\epsilon$ dimensions for this symmetry class in the presence of Coulomb interaction was developed in Refs~[\onlinecite{baranov99, baranov02}]. The dimensionless resistance $t$ is renormalized according to the following $\beta$-function:
\begin{equation}
-\frac{dt}{d\ln y} = \beta(t) = \epsilon t - t^2 - A t^3 + O(t^4) ,
\label{eq2E1}
\end{equation}
Here the numerical factor in the two-loop contribution is equal [\onlinecite{baranov02}]
\begin{gather}
A =\frac{1}{16}\Bigl [\frac{139}{6}+\frac{(\pi^2-18)^2}{12}+\frac{19}{2}\zeta (3)+\Bigl ( 16 + \frac{\pi ^2}{3} \Bigr )\ln ^{2}2  \notag \\
- \Bigl (44-\frac{\pi ^{2}}{2}+7\zeta (3)\Bigr ) \ln 2+16\mathcal{G}-\frac{1}{3}\ln ^{4}2-8\lit\left(\frac{1}{2}\right)\Bigr ]\notag \\ \approx 1.64 ,
\end{gather}
where $\mathcal{G} \approx 0.915 $ denotes the Catalan constant, $\zeta(x)$ stands for the Riemann zeta-function, and $\lit(x) = \sum_{k=1}^\infty x^k/k^4$ denotes the polylogarithm. As usual, the condition $\beta(t_*)=0$ determines the critical point:
$t_*=\epsilon(1-A\epsilon)+O(\epsilon^3)$ (and thus the critical
conductance $g_*=2/\pi t_*$). Then the multifractal exponents controlling scaling behavior of the moments of the local density of states read [\onlinecite{BGM2013}]
\begin{equation}
\Delta_q = \zeta_q^* =
\frac{q(1-q)\epsilon}{4}\Bigl [1 + \left (1-A-\frac{\pi^2}{12}\right ) \epsilon \Bigr ]+O(\epsilon^3) .
\label{eq2E3}
\end{equation}
The localization  length exponent is obtained as $\nu=-1/\beta^\prime(t_*)=1/\epsilon-A+O(\epsilon)$.
The dynamical exponent is also known up to the two-loop order:
$z=2+\epsilon/2+(2A-\pi^2/6-3)\epsilon^2/4+O(\epsilon^3)$ [\onlinecite{baranov99}].

For the case of broken time reversal and spin rotational symmetries, the anomalous dimension of the disorder-averaged local density of states becomes (cf. Eq. \eqref{eq:ad:Z})
\begin{equation}
\zeta = - \frac{1}{2} \ln(1+\gamma_s) t + O(t^2) .
\end{equation}
In the case of Coulomb interaction, $\gamma_s=-1$, one has to substitute $\ln(1+\gamma_s)$ by $-2/\epsilon$. It leads to the following results (see e.g., Refs. [\onlinecite{finkelstein90,belitz94}]):  
\begin{equation}
\theta = \frac{t_*}{\epsilon} = 1, \qquad \beta = \frac{\theta}{z} = \frac{1}{2} .
\end{equation}
Therefore, in $d=2+\epsilon$ the combination $\theta q +\Delta_q$ is positive for $q \lesssim 4/\epsilon$.

It is worthwhile to compare the results for interacting critical point with the known results for the
critical point in the absence of interactions. In the case of Anderson transition in the unitary Wigner-Dyson class A, the $\beta$-function, the critical point and the localization length exponent are
known up to the five-loop order [\onlinecite{HikamiWegnerB}]
\begin{equation}
-\frac{dt}{d\ln y} = \beta^{(0)}(t) = \epsilon t - \frac{1}{8} t^3 - \frac{3}{128} t^5 + O(t^6) ,
\label{eq2E4}
\end{equation}
$t_*=(2\epsilon)^{1/2}(1-3\epsilon/4)+O(\epsilon^{5/2})$, and
$\nu=1/2\epsilon-3/4+O(\epsilon)$. The anomalous dimensions of operators which
determine the scaling behavior of the moments of the local density of states have been computed at the
four-loop level  with the result [\onlinecite{Wegner}]
\begin{equation}
\zeta_q^{(0)}(t) = \frac{q(1-q) t}{4}\left ( 1+ \frac{3}{32}t^2 + \frac{3
\zeta(3)}{128} q(q-1) t^3\right )+ O(t^5) ,
\label{eq2E5}
\end{equation}
This leads to the following expression for the multifractal exponents:
\begin{equation}
\Delta_q^{(0)} = q(1-q) \left (\frac{\epsilon}{2}\right )^{1/2} - \frac{3 \zeta(3)}{32} q^2(q-1)^2 \epsilon^2  +  O(\epsilon^{5/2}) .
\label{eq2E6}
\end{equation}
Comparing Eqs. \eqref{eq2E1} and \eqref{eq2E4}, one sees that Coulomb interaction changes the $\beta$-function and, consequently, the fixed point and critical exponents. Thus, Anderson transitions with and without Coulomb interaction belong to  different universality classes. As a consequence, sets of mutlifractal exponents with and without Coulomb interaction are also different.

For small enough values of $\epsilon$, when the expansion in $t$ is parametrically
controlled, the Coulomb interaction considerably reduces numerical values of
the multifractal exponents, i.e. weakens multifractality. As an example, for
$\epsilon=1/9$ we get $\eta^{(0)} =-\Delta_2^{(0)}=0.48$ in the absence of interaction and ten times smaller value, $\eta =-\Delta_2=0.047$, in the presence of interaction. In the case of dimensionality $d=3$, i.e. $\epsilon=1$, we can only use our results as a rough estimate. Taking for this estimate the one-loop result we obtain $\eta^{(0)} \sim 1.4$ and $\eta \sim 0.5$, i.e.,
again the exponent $\eta$ for the interacting critical point is smaller than this exponent in the non-interacting case, $\eta^{(0)} > \eta$.
A qualitative reason for this is that in the considered symmetry class the interaction has a "localizing" effect (it suppresses the conductivity). Thus, the interaction shifts a critical point towards weaker disorder and thus weaker multifractality. While this is a controllable argument for Anderson transition at small $\epsilon$, for $d=3$ or higher dimensions (where the transition is not in the weak-coupling regime from the sigma model point of view) this is just a plausible (but not rigorous) reasoning.
 
In the case of the integer quantum Hall effect Eq. \eqref{eq2E1} (with $\epsilon=0$) describes the perturbative contributions to the renormalization of the resistivity along the line of half-integer Hall conductance. Although the perturbative result  \eqref{eq2E1} favors localization, there is a topological protection (nonperturbative in $t$ contributions to the beta-function [\onlinecite{burmistrov07}]) which leads to the existence of the critical point at some coupling $t\sim 1$. We thus expect that the multifractal exponents $\Delta_q$ at this critical point are of the order unity.  A similar scenario holds for the non-interacting electrons, see [\onlinecite{Evers08}] for a review. We do not know whether the quantum-Hall multifractality in the presence of interaction is stronger or weaker than in the non-interacting system (in particular, whether the interacting exponent, $\eta$, is smaller or larger than the non-interacting one, $\eta^{(0)}$, equal to $\approx 0.55$ according to numerical simulations).

\subsection{Broken spin rotational symmetry but preserved time reversal symmetry}
 \label{s4.4}

Now, we consider a system of disordered fermions  with Coulomb interaction in the presence of time-reversal symmetry but in the absence of spin rotational symmetry, which corresponds to the symmetry classes ``SO(LR)'' in terminology of Ref.~\onlinecite{belitz94}. In the absence of interactions, this situation corresponds to the symplectic Wigner-Dyson class AII. In this case, all triplet diffuson and cooperon modes ($W$ with $j=1,2,3$) are suppressed at large length scales. Then, Eqs. \eqref{eq1loopK2_3} and \eqref{eqK1_0} are transformed into the following two-loop result for the 2-point irreducible correlation function:
\begin{align}
K_2  & = -\rho_0^2 \frac{h^{\epsilon}t}{2\epsilon} + \rho_0^2 \frac{t^2\,h^{2\epsilon}}{4\epsilon^2} \Bigl [ 1- \Bigl (1 +\frac{\epsilon}{2}\Bigr ) + 2 \ln(1+\gamma_s)   \notag \\
&  + 2 f(\gamma_s) + \frac{\epsilon}{2} \bigl [ \ln(1+\gamma_s) + 2 f(\gamma_s) - c(\gamma_s)\bigr ]\Bigr ] . \label{eqK1_0-sym}
\end{align}
Hence, for the Coulomb interaction, $\gamma_s=-1$, the anomalous dimension of the $q$-th moment of the local density of states is given as follows 
\begin{equation}
\zeta_q = \frac{q(1-q)}{2} \Bigl [ \frac{t}{2} + c(-1)  \frac{t^2}{4} \Bigr ]+ O(t^3) .
\label{eqmqRG-sym}
\end{equation}
Here, we remind, the function $c(\gamma)$ is given in Eq. \eqref{eq:def:c} and $c(-1)=2-\pi^2/6$. In the case of the symmetry class ``SO(LR)'' the beta-function is known in the one-loop approximation only (see e.g. Refs. [\onlinecite{finkelstein90,belitz94}]): 
\begin{equation}
-\frac{dt}{d\ln y}  = \beta(t) =  \epsilon t - t^2 \Bigl [-\frac{1}{2} + f(-1) \Bigr ] +O(t^3) .
\label{eq2E1-sym}
\end{equation} 
Here the function $f(\gamma)$ is defined after Eq.
\eqref{eq2loopK2_P2+-2}. The contribution $-1/2$ is due to weak antilocalization whereas the term $f(-1)=1$
describes the Aronov-Altshuler contribution in the singlet channel which favors localization. To the lowest order in $\epsilon$ we find the following estimates for the critical dimensionless resistance and for the correlation length exponent: $t_*= 2 \epsilon$ and $\nu = 1/\epsilon$, respectively. Then, from Eq. \eqref{eqmqRG-sym}, we obtain  the following one-loop result 
\begin{equation}
\label{multifrac-exp-SOLR}
\Delta_q = q(1-q) \epsilon/2  
\end{equation}
for the multifractal exponents which governs behavior of the moments of the local density of states. We remind that for the symmetry class ``SO(LR)'' values of the dynamical exponent and the exponent of the disorder-averaged local density of states are $z = 2 + O(\epsilon^2)$ and $\theta=2$ ($\beta=1$), respectively (see e.g. Refs. [\onlinecite{finkelstein90,belitz94}]). 

In the absence of electron-electron interaction the one-loop renormalization group beta-function does not predict  Anderson transition in $d=2+\epsilon$ dimension: there is only one fixed point at $t=0$ corresponding to the metallic phase. This is also valid for a 2D system; in this case one finds a flow towards a supermetallic fixed point with an infinite conductivity. 
To recover the Anderson transition, one should take into account higher orders of the loop expansion [\onlinecite{HikamiWegnerB}]  and topological (vortex-like) excitations [\onlinecite{vortices}].  The Anderson (super-)metal-insulator transition in 2D and in $d=2+\epsilon$ dimensions is thus at strong coupling, i.e. at some $t \sim 1$, and cannot be studied analytically in a controllable way. In $d=2,3$ this transition was studied by numerical means [\onlinecite{Evers08}]. In view of the strong-coupling character of the non-interacting transition, the corresponding multifractal exponents $\Delta_q^{(0)}$ are of order unity in 2D and in $d=2+\epsilon$ dimensions. This should be contrasted to the above analysis of the interacting systems. The Coulomb interaction eliminates the 2D supermetallic phase, rendering the 2D system insulating. As a consequence, the Anderson transition in  $d=2+\epsilon$ dimensions acquires the weak-coupling character. In particular, the multifractal exponents are of order $\epsilon$ in the presence of Coulomb interaction, see Eq.\eqref{multifrac-exp-SOLR}.  Therefore, similar to the unitary case, the electron-electron interaction reduces the values of the exponents characterising the multifractality of the local density of states in the symplectic case in  $d=2+\epsilon$ dimensions (e.g., $\eta^{(0)} > \eta$). Again, it is plausible that such a reduction of exponents holds true also for 3D systems but we are not aware of any analytical argument proving this. Since the 3D transition is of strong-coupling character (whether with or without Coulomb interaction), numerical methods should be used to find precise numerical values of the multifractal exponents. 

In $d=2$ Eq. \eqref{eq2E1-sym} demonstrates tendency towards localization. However, in the case when the symmetry class ``SO(LR)'' corresponds to a single flavor of Dirac fermion, the localization is avoided due to a topological protection [\onlinecite{OGM2007}].  Such a situation is realized on the surface of a 3D $\mathbb{Z}_2$ topological insulator. The interacting system then flows into a fixed point with a coupling  $t\sim 1$ [\onlinecite{OGM2010}].  We thus expect that the corresponding multifractal exponents $\Delta_q$ are of the order of unity at this interacting critical point. This can be contrasted with the non-interacting case in which a model of a single flavor of Dirac fermion was numerically found to be always in the supermetallic phase  [\onlinecite{SM}], corresponding to $t=0$ in the infrared. 
Thus, the Coulomb interaction has a dramatic impact on properties of the system (including multifractality) in this situation. Specifically, it transforms the supermetallic phase (with no multifractality in the limit of large system) into a strong-coupling critical phase with strong multifractality (multifractal exponents of order unity).  

A similar situation occurs also in a transition between the normal and topological insulators in 2D [\onlinecite{OGM2010}]. Also in this problem, a strong-coupling fixed point emerges, instead of a supermetallic phase for a non-interacting system, due to an interplay of Coulomb interaction and topology (which is in this case implemented by vortices in the sigma-model language [\onlinecite{vortices}]). This interaction-induced fixed point can be also realised on a surface of a weak 3D topological insulator (cf. Ref.~[\onlinecite{ringel12}] where the corresponding non-interacting problem was analysed). As for other strong-coupling fixed points, we expect multifractal exponents of order unity in this problem. 

\subsection{Partially broken spin rotational symmetry \label{s4.5}}

The spin rotational symmetry can be broken not only due to spin-orbit coupling but also due to spin-orbit scattering. Provided the relaxation rates for $S_x$ and $S_y$ components of the spin due to spin-orbit scattering are much larger than for $S_z$, $1/\tau_z \ll 1/\tau_{x,y}$, the spin rotational symmetry is broken only partially. 
In this case the time reversal symmetry is preserved. Such situation can be expected for the 2D electrons with spin-orbit coupling and spin-orbit scattering (see e.g., Ref. [\onlinecite{finkelstein90}]. In this case the cooperon and diffuson modes with non-zero spin projections ($W$ with $j=1,2$) are suppressed.  Then, Eqs. \eqref{eq1loopK2_3} and \eqref{eqK1_0} transforms into
\begin{align}
K_2  & = -\rho_0^2 \frac{h^{\epsilon}t}{\epsilon} + \rho_0^2 \frac{t^2\,h^{2\epsilon}}{2\epsilon^2} \Biggl \{ 1 + \sum_{j=0,3} \Bigl [ 2 \ln(1+\gamma_j)   \notag \\
&  + 2 f(\gamma_j) + \frac{\epsilon}{2} \bigl [ \ln(1+\gamma_j) + 2 f(\gamma_j) - c(\gamma_j)\bigr ]\Bigl ] \Biggr \} . \label{eqK1_0-sym2}
\end{align}
This leads to the following result for the anomalous dimension of the $q$-th moment of the local density of states  
\begin{equation}
\zeta_q = \frac{q(1-q)}{2} \Bigl [ t + \bigl [c(\gamma_s)+c(\gamma_t)\bigr ] \frac{t^2}{2} \Bigr ]+ O(t^3) .
\label{eqmqRG-sym2}
\end{equation}

The renormalization group equations in this symmetry class are known up to the one-loop order only (see e.g. Ref. [\onlinecite{finkelstein90}]): 
\begin{align}
-\frac{dt}{d\ln y} & = \beta(t) =  \epsilon t - t^2 [f(\gamma_s) +  f(\gamma_t)] , \notag \\
-\frac{d\gamma_s}{dy} & = \frac{t}{2} (1+\gamma_s)(\gamma_s+\gamma_t) ,
\label{eq2E1-u3} \\
-\frac{d\gamma_t}{dy} & = \frac{t}{2} (1+\gamma_t)(\gamma_s+\gamma_t) . \notag 
\end{align}
In the case of Coulomb interaction, $\gamma_s=-1$, these renormalization group equations have a non-trivial interacting fixed point: $\gamma^*_t=1$ and $t_* = \epsilon/[2(1-\ln 2)]$. The corresponding one-loop results for the correlation length exponent, dynamical exponent and exponent of the disorder-averaged local density of states for this case are as follows [\onlinecite{belitz94}]: 
\begin{equation}
\nu = \frac{1}{\epsilon}, \qquad z=2+\epsilon, \qquad \theta= \frac{1}{2(1-\ln 2)} .
\label{set-exp-u3}
\end{equation}
Substituting the critical value  $t_*$ in the leading-order term of Eq.~(\ref{eqmqRG-sym2}), we obtain the multifractal exponents in the one-loop approximation:
\begin{equation}
\Delta_q = \frac{q(1-q)}{2(1-\ln 2)} \epsilon . 
\label{set-multi-sym2}
\end{equation}

We note that the same renormalization-group equations, Eqs. \eqref{eq2E1-u3}, describe also the case when a magnetic field breaks the time-reversal symmetry and partially destroys the spin invariance due to a finite value of the $g$-factor. In terminology of Ref. [\onlinecite{belitz94}] these are ``MF(LR)'' (for Coulomb interaction) and ``MF(SR)'' (for short-ranged interaction) classes. In this case, all cooperon modes ($W$ with $r=1,2$) and triplet diffuson modes with non-zero total spin projection ($W$ with $j=1,2$) are suppressed. Equations \eqref{eqmqRG-sym2}, \eqref{set-exp-u3}, and  \eqref{set-multi-sym2} are fully applicable to this situation as well.

\subsection{Preserved spin invariance\label{s4.3}}

Finally, we consider a system of disordered interacting fermions in the presence of spin rotation symmetry.  The key difference with the case of broken spin invariance is that now no weak-coupling transition point in $2+\epsilon$ dimensions is found. Instead, the system undergoes a metal-insulator transition at a coupling $t\sim 1$ in two dimensions.

We will focus here on a system with broken time-reversal invariance; the analysis of the time-reversal-invariant case leads to  similar results. The considered situation can be realized in the presence of magnetic field provided the $g$-factor is zero (see e.g., Ref. [\onlinecite{finkelstein90}]). In the case of finite small value of $g$-factor, spin rotational symmetry is preserved at the intermediate length scales. In the absence of electron-electron interactions this corresponds to  the two independent unitary Wigner-Dyson classes A for each spin component. In this case, all cooperon modes ($W$ with $r=1,2$) are suppressed. Then, Eqs. \eqref{eq1loopK2_2}and \eqref{eqK1_0} transform into the following result:
\begin{align}
K_2  & = -\rho_0^2 \frac{h^{\epsilon}t}{\epsilon} + \rho_0^2 \frac{t^2\,h^{2\epsilon}}{2\epsilon^2} \Biggl \{ 1 + \sum_{j=0}^3 \Bigl [ 2 \ln(1+\gamma_j)   \notag \\
&  + 2 f(\gamma_j) + \frac{\epsilon}{2} \bigl [ \ln(1+\gamma_j) + 2 f(\gamma_j) - c(\gamma_j)\bigr ]\Bigl ] \Biggr \} . \label{eqK1_0-u2}
\end{align}
Hence, the anomalous dimension of the $q$-th moment of the local density of states is given as follows: 
\begin{equation}
\zeta_q = \frac{q(1-q)}{2} \Bigl [ t + \bigl [c(\gamma_s)+3c(\gamma_t)\bigr ] \frac{t^2}{2} \Bigr ]+ O(t^3) .
\label{eqmqRG-u2}
\end{equation}
We remind that the function $c(\gamma)$ is given in Eq. \eqref{eq:def:c}. The renormalization group equations are known up to the one-loop approximation only (see e.g. Refs. [\onlinecite{finkelstein90,belitz94}]): 
\begin{align}
-\frac{dt}{d\ln y} & =   \epsilon t - t^2 [f(\gamma_s) + 3 f(\gamma_t)] , \notag \\
-\frac{d\gamma_s}{dy} & = \frac{t}{2} (1+\gamma_s)(\gamma_s+3\gamma_t) ,
\label{eq2E1-u2} \\
-\frac{d\gamma_t}{dy} & = \frac{t}{2} (1+\gamma_t)(\gamma_s-\gamma_t) . \notag 
\end{align}

We note that the one-loop result for $\zeta^{(0)}_q$ (which coincides with the one-loop result in the presence of interaction) can be obtained from renormalization group equations \eqref{eq2E1-u2} expanded to the lowest order in interaction amplitudes $\gamma_s$ and $\gamma_t$:
\begin{equation}
\frac{d}{dy} \begin{pmatrix}
\gamma_s\\
\gamma_t
\end{pmatrix} = -\frac{t}{2} \mathcal{R}  \begin{pmatrix}
\gamma_s\\
\gamma_t
\end{pmatrix}, \quad \mathcal{R}  = \begin{pmatrix}
1 & 3\\
1 & -1 
\end{pmatrix} .
\end{equation}
The matrix $\mathcal{R}$ has two eigenvalues of opposite sign: $\lambda_\pm = \pm 2$. Then the one-loop result for the anomalous dimension in the non-interacting case can be written as follows: $\zeta_q^{(0)} = \lambda_- q(q-1) t/4$. 

As is well known [\onlinecite{finkelstein90,belitz94}], the one-loop renormalization group equations \eqref{eq2E1-u2} for the case of preserved spin invariance are not sufficient to describe the Anderson transition. The reason for this is a growth of the triplet-channel coupling  $\gamma_t$ which eventually forces the resistivity $t$ to flow towards a metal. One thus needs a two-loop extension of these equations.  This has been achieved in the limit of large number $n_v$ of valleys (fermion flavors) [\onlinecite{punnoose05}].  The one-loop renormalization group equations \eqref{eq2E1-u2} in two dimensions are modified in the case of arbitrary $n_v$ as follows:
\begin{align}
-\frac{dt}{d\ln y} & =    - \frac{t^2}{n_v} \bigl [f(-1) + (4n_v^2-1) f(\gamma_t/n_v) \bigr ] , \notag \\
-\frac{d\gamma_t}{dy} & = - \frac{t}{2} \left (1+\frac{\gamma_t}{n_v}\right )^2 . \label{eq2E1-u2-nv} 
\end{align}
where $t$ and $\gamma_t$ are resistivity and triplet-channel interaction per single fermion flavor, respectively. It was found in Ref. [\onlinecite{punnoose05}] that within the two-loop generalisation of the renormalization group equations (\ref{eq2E1-u2-nv}) there exists a fixed point describing the metal-insulator transition at $t=t_*\approx 0.3$ and $\gamma_t=\gamma_{t}^* \approx 1.5$. 

The anomalous dimension \eqref{eqmqRG-u2} of the $q$-th moment of the local density of states in the case of an arbitrary $n_v$ becomes
\begin{equation}
\zeta_q = \frac{q(1-q)}{2} \Bigl [ t + \bigl [c(-1)+(4n_v^2-1)c(\gamma_t/n_v)\bigr ] \frac{t^2}{2n_v} \Bigr ]+ O(t^3) .
\label{eqmqRG-u2-nv}
\end{equation}
Using the result \eqref{eqmqRG-u2-nv} in the large-$n_v$ limit as well as the above value of $t_*$, we obtain a two-loop approximation for the multifractal exponents 
\begin{equation}
\Delta_q = \frac{q(1-q)}{2} \eta, \qquad \eta = t_* \approx -0.3 .
\label{eqmqRG-u2-nv-l}
\end{equation}
We note that  the two-loop approximation is not exact (even in the limit $n_v\to \infty$). Therefore, this result for the multifractal spectrum at the 2D metal-insulator transition point should be considered as a rough estimate only.

\section{Conclusions}
\label{s5}

In this paper we have shown that the multifractal fluctuations and correlations
of the local density of states persist in the presence of Coulomb interaction. By using the
non-linear sigma-model approach, we have calculated the multifractality spectrum for interacting systems
with different symmetries (with respect to time reversal and spin rotations) up to the two-loop
order in $2+\epsilon$ dimensions. For systems with fully preserved spin-rotation invariance our analysis yields an estimate for the multufractality spectrum at the 2D metal-insulator transition. For all symmetry classes, the obtained values of the multifractal exponents are essentially different from their non-interacting counterparts. This happens both because of a difference in the corresponding anomalous scaling functions and because of different values of the critical resistance $t_*$. We mention that in all cases the spectrum of multifractal dimensions $\Delta_q$ (and thus the singularity
spectrum $f(\alpha)$ that is obtained by the Legendre transformation, see
Ref. [\onlinecite{Evers08}]) is parabolic, $\Delta_q \simeq \gamma q(1-q)$, in the two-loop approximation. It is expected, however, that
higher-loop contributions will break the exact parabolicity in the case of Coulomb interaction, in analogy with what happens (in the four-loop order) in the non-interacting model [\onlinecite{Wegner}].

We hope that our work will motivate further experimental and numerical studies of
multifractality of interacting electrons near metal-insulator transitions, quantum Hall plateau-plateau transitions
and transitions between different phases of topological insulators. On the theoretical side, our paper paves a way to a systematic investigation of multifractality at interacting critical points of localization
transitions within the nonlinear sigma model approach. The rich physics related to
multifractality in the absence of interaction, including, in particular, systems
of different symmetry classes and different dimensionalities, symmetries of
mutlifractal spectra, termination and freezing, implications of
conformal symmetry, connection to entanglement entropy, and manifestation of
multifractality in various observables
[\onlinecite{Evers08,jia08,obuse10,gruzberg11,gruzberg13}] remains to be explored in the presence of Coulomb
interaction. Finally, we mention that our analysis of multifractal correlations in the local density of states can be extended to superconductor-insulator transitions [\onlinecite{Future}].

\begin{acknowledgments}

We thank A.W.W.~Ludwig, V.~Kravtsov, M. M\"uller, B.I.~Shklovskii, M.~Skvortsov, A.~Yazdani for discussions.
The work was supported by the program DFG SPP 1666 ''Topological insulators'', German-Israeli Foundation, Dynasty Foundation, RFBR grant No. 14-02-00333, and Russian Ministry of Education and Science under grant No. 14Y.26.31.0007.

\end{acknowledgments}

\appendix

\section{Evaluation of the two-loop integrals in $d=2+\epsilon$ dimension}
\label{app:sec:1}

In this appendix we present results for the two-loop integrals which determine the infra-red behavior of  
$[P_2^{\alpha_1\alpha_2}]^{RR (2)}(E,E^\prime)$ and $ [P_2^{\alpha_1\alpha_2}]^{RA (2)}(E,E^\prime)$. Following, we do not distinguish between diffuson and cooperon propagators denoting both by $\mathcal{D}$. Then a general integral that we need to consider is as follows:
\begin{equation}
J^{\delta}_{\nu\mu\eta}(\alpha) = \int_{pq}\int_0^\infty d s\,
s^\delta
\mathcal{D}^\nu_p(s)\mathcal{D}^a_p(s)\mathcal{D}_q^\mu(0)D_{\bm{p}+\bm{q}}^\eta(s) ,
\end{equation}
where $\mathcal{D}^a_p(s) = [p^2+h^2+a s]^{-1}$. In the limit $E=E^\prime=T=0$ we find
from Eqs. \eqref{eq2loopK2_P2++11} and \eqref{eq2loopK2_P2+-1}
\begin{equation}
[P_2^{\alpha_1\alpha_2}]^{RR (2)} \to 2 \left (\frac{16}{g}\right )^2 \sum_{j=0}^3 \gamma_j J^0_{101}(1+\gamma_j) ,
\label{app1:prr}
\end{equation}
and
\begin{gather}
[P_2^{\alpha_1\alpha_2}]^{RA (2)} \to  -2 \left (\frac{16}{g}\right )^2 \Biggl \{  \sum_{j=0}^3 \gamma_j \Bigl [ J^0_{110}(1+\gamma_j) \notag \\
+ J^0_{020}(1+\gamma_j) - J^0_{020}(1) + \gamma_j J^1_{021}(1+\gamma_j)
\Bigr ] \notag \\
+ \int_{qp} \mathcal{D}_q(0) \mathcal{D}_p(0) \bigl [ 3 + 2 p^2 \mathcal{D}_q(0) \bigr ]  \Biggr \}.
\label{app1:pra}
\end{gather}

Using the result
\begin{equation}
\int_q \mathcal{D}_q(0) = -\frac{2\Omega_d h^{\epsilon}
\Gamma(1+\epsilon/2)\Gamma(1-\epsilon/2)}{\epsilon} , \label{app1:Int1}
\end{equation}
we find
\begin{gather}
\int_{qp} \mathcal{D}_q(0) \mathcal{D}_p(0) = \frac{4A_\epsilon h^{2\epsilon}}{\epsilon^2}
, \notag \\
\int_{qp} p^2 \mathcal{D}^2_q(0) \mathcal{D}_p(0) = \frac{2A_\epsilon h^{2\epsilon}}{\epsilon} ,
\label{J0}
\end{gather}
where $A_\epsilon = \Omega_d^2 \Gamma^2(1-\epsilon/2)\Gamma^2(1+\epsilon/2)$. Next, with the help of the result
\begin{equation}
\int_0^\infty ds \int_q \mathcal{D}_q^a(s) = \frac{4 \Omega_d
h^{\epsilon+2}
\Gamma(1+\epsilon/2)\Gamma(1-\epsilon/2)}{a\epsilon(2+\epsilon)} 
\end{equation}
we obtain
\begin{equation}
J^0_{020}(a) = \frac{4A_\epsilon
h^{2\epsilon}}{a\epsilon(2+\epsilon)} . \label{J0_020}
\end{equation}
Using the relation
\begin{equation}
\int_0^\infty d\omega \int_q \mathcal{D}_q(\omega)\mathcal{D}^a_q(\omega) = \frac{\ln a }{a-1}
\int_q \mathcal{D}_q(0)
\end{equation}
we derive
\begin{equation}
J^0_{110}(a) = \frac{4 A_\epsilon h^{2\epsilon}}{\epsilon^2}
\frac{\ln a}{a-1} . \label{J0_110}
\end{equation}

It is convenient to rewrite the integral $J^0_{101}(a)$ as follows
\begin{equation}
J^0_{101}(a) = -\frac{A_\epsilon \Gamma(1-\epsilon)
h^{2\epsilon}}{\epsilon \Gamma^2(1-\epsilon/2)} T_{01}(a)
\end{equation}
where (see Eq.(A26) of Ref. [\onlinecite{baranov02}])
\begin{gather}
T_{01}(a)= \int_0^1 dx_1 \int_0^1 dx_2  \int_0^1 dx_3 \, \delta(x_1+x_2+x_3-1)  \notag \\
\times \frac{x^{-1-\epsilon/2}_3
(x_1+x_2)^{-1-\epsilon/2}}{x_{1}+a x_2 + x_3} .
\end{gather}
The evaluation yields
\begin{gather}
T_{01}(a)=\frac{2\Gamma^2(1-\epsilon/2)}{(a-1)\epsilon
\Gamma(1-\epsilon)} \int_{a}^{1}
\frac{du}{u^{1+\epsilon/2}}\notag \\
\times {}_{2}F_{1}(-\epsilon/2,-\epsilon,1-\epsilon,1-u)
= -\frac{\Gamma^2(1-\epsilon/2)}{(a-1)\Gamma(1-\epsilon)}\notag \\
\times \left [
\frac{2 \ln a}{\epsilon} -\frac{1}{2}\ln^2a\right ] .
\end{gather}
Hence,
\begin{equation}
J^0_{101}(a) = \frac{A_\epsilon h^{2\epsilon}}{a-1} \left
[\frac{2\ln a}{\epsilon^2} -\frac{\ln^2a}{2\epsilon} \right ] .
\label{J0_101}
\end{equation}

We rewrite the integral $J^1_{021}(a)$ in the following way:
\begin{equation}
J^1_{021}(a)=-\frac{A_\epsilon \Gamma(1-\epsilon)
h^{2\epsilon}}{\epsilon \Gamma^2(1-\epsilon/2)} S^{1}_{2}(a) ,
\end{equation}
where 
\begin{gather}
S^1_2(a)=\int_0^1 dx_1 \int_0^1 dx_2  \int_0^1 dx_3 \, \delta(x_1+x_2+x_3-1) \notag \\ 
(x_1x_2+x_2x_3+x_3x_1)^{-1-\epsilon/2} \frac{x_2}{ (a
x_{1}+x_3)^2} .
\end{gather}
Evaluating the integral, we find
\begin{widetext}
\begin{gather}
S^1_2(a) = \frac{4}{\epsilon(2+\epsilon)} \int_0^1 \frac{du\,
[u(1-u)]^{1-\epsilon}}{(au+1-u)^2} {}_{2}F_{1}(2,-\epsilon,1-\epsilon/2,1-u(1-u)) 
= \frac{2}{\epsilon} \frac{2(1-a)+(1+a)\ln a}{(a-1)^3}\notag  \\
-\frac{2}{a} + \frac{2(1+a)\ln a}{(a-1)^3} +\frac{2(1+a)}{(a-1)^3}
\left [\liq(1-a)+\frac{1}{4}\ln^2 a\right ] +O(\epsilon) .
\end{gather}
Hence,
\begin{equation}
J^1_{021}(a) = A_\epsilon h^{2\epsilon}\left [
- 2 \frac{2(1-a)+(1+a)\ln a}{(a-1)^3\epsilon^2} +\frac{2}{a\epsilon}
- \frac{2(1+a)\ln a}{(a-1)^3\epsilon} -\frac{2(1+a)}{(a-1)^3\epsilon}
\left [\liq(1-a)+\frac{1}{4}\ln^2 a\right ] + O(\epsilon)
  \right  ] .\label{J1_021}
\end{equation}

\section{One-loop renormalization of diffuson and cooperon propagators}
\label{app:sec:2}

In this appendix we demonstrate that the most part of the two-loop contribution to $[P_2^{\alpha_1\alpha_2}]^{RA}(E,E^\prime)$ can be considered as the renormalization of the diffuson and cooperon which determine one-loop contribution to  $[P_2^{\alpha_1\alpha_2}]^{RA}(E,E^\prime)$.

Taking into account Eq. \eqref{eq18P1} and Eq. \eqref{eq2loopK2_P2+-1}, we can rewrite expression for $[P_2^{\alpha_1\alpha_2}]^{RA}(E,E^\prime)$ in the following way
\begin{gather}
[P_2^{\alpha_1\alpha_2}]^{RA}(E,E^\prime) = - 256 \int_q \frac{Z(E,E^\prime)}{g q^2 - 16 i z \Omega - \Sigma^R(q,E,E^\prime)} 
- 2 \left (\frac{16}{g} \right )^2 \left (\int_q \mathcal{D}^R_q(\Omega) \right )^2  
- 4 \left (\frac{16}{g} \right )^2 \int_{qp} \mathcal{D}^{R2}_q(\Omega)  .
\end{gather}
Here the renormalization factor $Z(E,E^\prime)$ is given as (cf. Eq. \eqref{eq_Z})
\begin{gather}
Z(E,E^\prime) = 1 + \frac{16}{i g^2} \sum_{j=0}^3 \Gamma_j \int_{p,\omega} \Bigl [ \mathcal{F}_{\omega-E}+  \mathcal{F}_{\omega+E^\prime}\Bigr ] 
\mathcal{D}^{R}_p(\omega)\mathcal{D}^{(j)R}_p(\omega) .
\end{gather}
The diffuson self-energy reads
\begin{gather}
\Sigma^R(q,E,E^\prime)   = 4 q^2 \int_p \mathcal{D}^{R}_p(\Omega)
- \frac{8}{ig} \sum_{j=0}^3 \Gamma_j \int_{p,\omega}
\Bigl [2\mathcal{B}_\omega- \mathcal{F}_{\omega-E}-  \mathcal{F}_{\omega+E^\prime}\Bigr ]   \frac{\mathcal{D}^{(j)R}_p(\omega)}{\mathcal{D}^{R}_p(\omega)} 
\Bigl [ \mathcal{D}^{R}_{\bm{p}+\bm{q}}(\omega+\Omega) + \mathcal{D}^{A}_{\bm{p}-\bm{q}}(\omega-\Omega) \Bigr ]
\notag \\
 + \frac{8}{ig} \sum_{j=0}^3 \Gamma_j \int_{p,\omega} \Bigl [ \mathcal{F}_{\omega-E}+  \mathcal{F}_{\omega+E^\prime}\Bigr ]\mathcal{D}^{(j)R}_p(\omega) \Bigl [ 2 \bm{p} \bm{q} \mathcal{D}^{R}_{\bm{p}+\bm{q}}(\omega+\Omega) + \bigl [\mathcal{D}^{R}_q(\Omega)\bigr ]^{-1}
\bigl[ \mathcal{D}^{R}_{\bm{p}+\bm{q}}(\omega+\Omega) - \mathcal{D}^{R}_p(\omega)\bigr ]
\Bigr ] .
\end{gather}
Expanding the self-energy $\Sigma^R(q,E,E^\prime)$ to the lowest order in $\omega$ and $q^2$, we find 
\begin{equation}
\frac{1}{g} \mathcal{D}^R_q(\Omega) \quad  \to \quad \frac{Z(E,E)}{g(E) q^2 - i 16 z(E) \Omega+\tau_\phi^{-1}(E)} .
\end{equation}
Here, we obtain
\begin{align}
g(E) & = g - 4 \int_p \mathcal{D}_p ^R(0)+ \frac{16}{g} \sum_{j=0}^3  \Gamma_j \int_{p,\omega}p^2
 \Bigl [\mathcal{F}_{\omega-E}+  \mathcal{F}_{\omega+E}\Bigr ] \im \Bigl [ \mathcal{D}^{(j) R}_p(\omega) \mathcal{D}^{R2}_{p}(\omega)\Bigr ] \notag \\
 & -\frac{16}{g} \sum_{j=0}^3  \Gamma_j \int_{p,\omega}
 \Bigl [2 \mathcal{B}_\omega - \mathcal{F}_{\omega-E}-  \mathcal{F}_{\omega+E}\Bigr ] \im \Bigl [ \mathcal{D}^{(j) R}_p(\omega) [\mathcal{D}^{R}_{p}(\omega)]^{-1}\Bigr ] \re \Bigr [ [1-2p^2\mathcal{D}^{R}_{p}(\omega)] \mathcal{D}^{R2}_{p}(\omega)\Bigr ] , \label{eqgE} \\
z(E) & = z + \frac{1}{2 g} \sum_{j=0}^3 \Gamma_j  \int_{p,\omega} \partial_\omega\Bigl [\mathcal{F}_{\omega-E}+  \mathcal{F}_{\omega+E}\Bigr ] \re \Bigl [ \mathcal{D}^{(j)R}_p(\omega) [\mathcal{D}^R_{p}(\omega)]^{-1} \Bigr ] \re \mathcal{D}^R_{p}(\omega) , \label{eqzE} \\
\tau_\phi^{-1}(E) & =\frac{4}{g} \sum_{j=0}^3 \Gamma_j  \int_{p\omega}\Bigl [2\mathcal{B}_\omega- \mathcal{F}_{\omega-E}-  \mathcal{F}_{\omega+E}\Bigr ]  \im \Bigl [ \mathcal{D}^{(j)R}_p(\omega) [\mathcal{D}^R_{p}(\omega)]^{-1} \Bigr ] \re \mathcal{D}^R_{p}(\omega) . \label{eqTE}
\end{align}
As one can see, indeed for $|E|\gg T$, the energy $|E|$ serves as cut-off for the infrared logarithmic divergences in Eqs. \eqref{eqgE} and \eqref{eqzE} (in the case of $L\to \infty$). At the same time non-zero value of energy $E$ induces non-zero dephasing time.

\section{The third and forth moment of the local density of states}
\label{app:sec:3}

In this appendix we calculate the two-loop contributions to the irreducible third and forth moments of the local density of states and demonstrate validity of Eq. \eqref{eqKmq}.

\subsection{The third irreducible moment of the local density of states}

The third irreducible moment $K_3$ can be obtained from the function
 \begin{align}
 K_3 &= \left (\frac{\rho_0}{8}\right )^3 \Bigl (P_3^{\alpha_1\alpha_2\alpha_3}(i\varepsilon_{n_1},i\varepsilon_{n_3},i\varepsilon_{n_5}) - P^{\alpha_1\alpha_2\alpha_3}_3(i\varepsilon_{n_1},i\varepsilon_{n_3},i\varepsilon_{n_2}) - P^{\alpha_1\alpha_2\alpha_3}_3(i\varepsilon_{n_1},i\varepsilon_{n_2},i\varepsilon_{n_3})
 - P^{\alpha_1\alpha_2\alpha_3}_3(i\varepsilon_{n_2},i\varepsilon_{n_1},i\varepsilon_{n_3})\notag \\
  & + P^{\alpha_1\alpha_2\alpha_3}_3(i\varepsilon_{n_2},i\varepsilon_{n_4},i\varepsilon_{n_1}) + P^{\alpha_1\alpha_2\alpha_3}_3(i\varepsilon_{n_2},i\varepsilon_{n_1},i\varepsilon_{n_4})
 + P^{\alpha_1\alpha_2\alpha_3}_3(i\varepsilon_{n_1},i\varepsilon_{n_2},i\varepsilon_{n_4})-P^{\alpha_1\alpha_2\alpha_3}_3(i\varepsilon_{n_2},i\varepsilon_{n_4},i\varepsilon_{n_6})
 \Bigl ) 
  \end{align}
 after analytic continuation to the real frequencies: $\varepsilon_{n_{1,3,5}} \to E+i0^+$ and $\varepsilon_{n_{2,4,6}} \to E-i0^+$. Here,
 \begin{align}
 P_3^{\alpha_1\alpha_2\alpha_3}(i\varepsilon_{n}, i\varepsilon_{m}, i\varepsilon_{k}) & = \langle \spp Q_{nn}^{\alpha_1\alpha_1}(\bm{r})  \spp Q_{mm}^{\alpha_2\alpha_2}(\bm{r})\spp Q_{kk}^{\alpha_3\alpha_3}(\bm{r}) \rangle - 3 \langle \spp Q_{nn}^{\alpha_1\alpha_1}(\bm{r}) \rangle \langle  \spp Q_{mm}^{\alpha_2\alpha_2}(\bm{r}) \spp Q_{kk}^{\alpha_3\alpha_3}(\bm{r}) \rangle 
 \notag \\
 & -3 \langle \spp Q_{nn}^{\alpha_1\alpha_1}(\bm{r}) \spp [ Q_{mk}^{\alpha_2\alpha_3}(\bm{r}) Q_{km}^{\alpha_3\alpha_2}(\bm{r}) ] \rangle +6 \langle \spp Q_{nn}^{\alpha_1\alpha_1}(\bm{r})\rangle \spp \langle Q_{mk}^{\alpha_2\alpha_3}(\bm{r}) Q_{km}^{\alpha_3\alpha_2}(\bm{r}) \rangle
 \notag \\
 &
 +8\spp  \langle Q_{nm}^{\alpha_1\alpha_2}(\bm{r}) Q_{mk}^{\alpha_2\alpha_3}(\bm{r}) Q_{kn}^{\alpha_3\alpha_1}(\bm{r}) \rangle +2 \langle \spp Q_{nn}^{\alpha_1\alpha_1}(\bm{r}) \rangle \langle \spp Q_{mm}^{\alpha_2\alpha_2}(\bm{r}) \rangle \langle  \spp Q_{kk}^{\alpha_3\alpha_3}(\bm{r}) \rangle
 \end{align}
 and replica indices $\alpha_1$, $\alpha_2$ and $\alpha_3$ are all different. In the two-loop approximation we find
 \begin{gather}
 P_3^{\alpha_1\alpha_2\alpha_3}(i\varepsilon_{n_1},i\varepsilon_{n_3},i\varepsilon_{n_2}) = P^{\alpha_1\alpha_2\alpha_3}_3(i\varepsilon_{n_1},i\varepsilon_{n_2},i\varepsilon_{n_3}) = P^{\alpha_1\alpha_2\alpha_3}_3(i\varepsilon_{n_2},i\varepsilon_{n_1},i\varepsilon_{n_3})= 
 - \left ( \frac{128}{g}\right )^2 \int_{qp} \mathcal{D}_q(i\Omega_{12}^\varepsilon) \mathcal{D}_p(i\Omega_{32}^\varepsilon) ,
 \notag\\
 P^{\alpha_1\alpha_2\alpha_3}_3(i\varepsilon_{n_2},i\varepsilon_{n_4},i\varepsilon_{n_1}) = P^{\alpha_1\alpha_2\alpha_3}_3(i\varepsilon_{n_2},i\varepsilon_{n_1},i\varepsilon_{n_4})
 = P^{\alpha_1\alpha_2\alpha_3}_3(i\varepsilon_{n_1},i\varepsilon_{n_2},i\varepsilon_{n_4}) = \left ( \frac{128}{g}\right )^2 \int_{qp} \mathcal{D}_q(i\Omega_{12}^\varepsilon) \mathcal{D}_p(i\Omega_{14}^\varepsilon)  ,
 \notag \\
 P^{\alpha_1\alpha_2\alpha_3}_3(i\varepsilon_{n_1},i\varepsilon_{n_3},i\varepsilon_{n_5}) = P^{\alpha_1\alpha_2\alpha_3}_3(i\varepsilon_{n_2},i\varepsilon_{n_4},i\varepsilon_{n_6}) = 0 .
 \end{gather}
Hence, we obtain
\begin{equation}
K_3 = \rho_0^3 \frac{12 t^2 h^{2\epsilon}}{\epsilon^2} +O(1) .
\end{equation}
By using Eq. \eqref{eqmqe3}, we obtain Eq. \eqref{eqmqe}.

\subsection{The $4$-th irreducible moment of the local density of states}

The $4$-th irreducible moment $K_4$ can be obtained from the function
 \begin{align}
 K_4 &= \left (\frac{\rho_0^4}{8}\right )^4 \Bigl (P^{\alpha_1\alpha_2\alpha_3\alpha_4}_4(i\varepsilon_{n_1},i\varepsilon_{n_3},i\varepsilon_{n_5},i\varepsilon_{n_7}) - P^{\alpha_1\alpha_2\alpha_3\alpha_4}_4(i\varepsilon_{n_1},i\varepsilon_{n_3},i\varepsilon_{n_5},i\varepsilon_{n_2}) 
 - P^{\alpha_1\alpha_2\alpha_3\alpha_4}_4(i\varepsilon_{n_1},i\varepsilon_{n_3},i\varepsilon_{n_2},i\varepsilon_{n_5})
  \notag \\
&
 - P^{\alpha_1\alpha_2\alpha_3\alpha_4}_4(i\varepsilon_{n_1},i\varepsilon_{n_2},i\varepsilon_{n_3},i\varepsilon_{n_5}) 
 - P^{\alpha_1\alpha_2\alpha_3\alpha_4}_4(i\varepsilon_{n_2},i\varepsilon_{n_1},i\varepsilon_{n_3},i\varepsilon_{n_5})  
 +P^{\alpha_1\alpha_2\alpha_3\alpha_4}_4(i\varepsilon_{n_1},i\varepsilon_{n_3},i\varepsilon_{n_2},i\varepsilon_{n_4}) 
   \notag \\
&
 +P^{\alpha_1\alpha_2\alpha_3\alpha_4}_4(i\varepsilon_{n_1},i\varepsilon_{n_2},i\varepsilon_{n_3},i\varepsilon_{n_4}) 
  +P^{\alpha_1\alpha_2\alpha_3\alpha_4}_4(i\varepsilon_{n_2},i\varepsilon_{n_1},i\varepsilon_{n_3},i\varepsilon_{n_4}) 
 +P^{\alpha_1\alpha_2\alpha_3\alpha_4}_4(i\varepsilon_{n_2},i\varepsilon_{n_4},i\varepsilon_{n_1},i\varepsilon_{n_3}) 
 \notag \\
&
+P^{\alpha_1\alpha_2\alpha_3\alpha_4}_4(i\varepsilon_{n_2},i\varepsilon_{n_1},i\varepsilon_{n_4},i\varepsilon_{n_3})
 +P^{\alpha_1\alpha_2\alpha_3\alpha_4}_4(i\varepsilon_{n_1},i\varepsilon_{n_2},i\varepsilon_{n_4},i\varepsilon_{n_3})
  - P^{\alpha_1\alpha_2\alpha_3\alpha_4}_4(i\varepsilon_{n_2},i\varepsilon_{n_4},i\varepsilon_{n_6},i\varepsilon_{n_1}) 
 \notag \\
&
 - P^{\alpha_1\alpha_2\alpha_3\alpha_4}_4(i\varepsilon_{n_2},i\varepsilon_{n_4},i\varepsilon_{n_1},i\varepsilon_{n_6})
 - P^{\alpha_1\alpha_2\alpha_3\alpha_4}_4(i\varepsilon_{n_2},i\varepsilon_{n_1},i\varepsilon_{n_4},i
 \varepsilon_{n_6}) 
 - P^{\alpha_1\alpha_2\alpha_3\alpha_4}_4(i\varepsilon_{n_1},i\varepsilon_{n_2},i\varepsilon_{n_4},i\varepsilon_{n_6})
   \notag \\
&
 +P^{\alpha_1\alpha_2\alpha_3\alpha_4}_4(i\varepsilon_{n_2},i\varepsilon_{n_4},i\varepsilon_{n_6},i\varepsilon_{n_8})
   \Bigl )  \label{eqK4P4}
 \end{align}
  after analytic continuation to the real frequencies: $\varepsilon_{n_{1,3,5,7}} \to E+i0^+$ and $\varepsilon_{n_{2,4,6,8}} \to E-i0^+$. Here,
 \begin{align}
 P_4^{\alpha_1\alpha_2\alpha_3\alpha_4} (i\varepsilon_{n}, i\varepsilon_{m}, i\varepsilon_{k} ,i\varepsilon_{l})
   & = \langle \spp Q_{nn}^{\alpha_1\alpha_1} \spp Q_{mm}^{\alpha_2\alpha_2} \spp Q_{kk}^{\alpha_3\alpha_3} \spp Q_{ll}^{\alpha_4\alpha_4} \rangle 
  -12 \langle \spp Q_{nn}^{\alpha_1\alpha_1} \spp Q_{ll}^{\alpha_4\alpha_4} \spp[ Q_{mk}^{\alpha_1\alpha_2} Q_{km}^{\alpha_2\alpha_1}] \rangle  \notag \\
 & - 32 \langle \spp Q_{nn}^{\alpha_1\alpha_1} \spp [ Q_{lm}^{\alpha_4\alpha_2} Q_{mk}^{\alpha_2\alpha_3} Q_{kl}^{\alpha_3\alpha_4} ] \rangle
   -48 \spp \langle Q_{nm}^{\alpha_1\alpha_2} Q_{mk}^{\alpha_2\alpha_3} Q_{kl}^{\alpha_3\alpha_4}
 Q_{ln}^{\alpha_4\alpha_1} \rangle  \notag \\
 & + 12 \langle \spp [ Q_{nm}^{\alpha_1\alpha_2}Q_{mn}^{\alpha_2\alpha_1}] \spp[ Q_{kl}^{\alpha_3\alpha_4}
Q_{lk}^{\alpha_4\alpha_3} ] \rangle 
      - 4 \langle \spp Q_{nn}^{\alpha_1\alpha_1} \rangle \langle  \spp Q_{mm}^{\alpha_2\alpha_2} \spp Q_{kk}^{\alpha_3\alpha_3} \spp Q_{ll}^{\alpha_4\alpha_4} \rangle 
      \notag \\
 & +12 \langle \spp Q_{nn}^{\alpha_1\alpha_1} \rangle \langle \spp Q_{ll}^{\alpha_4\alpha_4} \spp[ Q_{mk}^{\alpha_2\alpha_3} Q_{km}^{\alpha_3\alpha_2}] \rangle 
-32 \langle \spp Q_{nn}^{\alpha_1\alpha_1} \rangle \langle \spp [Q_{lm}^{\alpha_4\alpha_2} Q_{mk}^{\alpha_2\alpha_3} Q_{kl}^{\alpha_3\alpha_4} ] \rangle\notag \\
& +6 \langle \spp Q_{nn}^{\alpha_1\alpha_1}\rangle  \langle \spp Q_{ll}^{\alpha_4\alpha_4} \rangle \langle  \spp Q_{mm}^{\alpha_2\alpha_2} \spp Q_{kk}^{\alpha_3\alpha_3} \rangle
-12 \langle \spp Q_{nn}^{\alpha_1\alpha_1}\rangle  \langle \spp Q_{ll}^{\alpha_4\alpha_4} \rangle \langle  \spp [Q_{mk}^{\alpha_2\alpha_3} Q_{km}^{\alpha_3\alpha_2} \rangle
\notag \\
   &  -3  \langle \spp Q_{nn}^{\alpha_1\alpha_1} \rangle \langle \spp Q_{mm}^{\alpha_2\alpha_2} \rangle \langle  \spp Q_{kk}^{\alpha_3\alpha_3} \rangle
\langle  \spp Q_{ll}^{\alpha_4\alpha_4} \rangle 
 \end{align}
 and replica indices $\alpha_1$, $\alpha_2$, $\alpha_3$, $\alpha_4$ are all different.
  In the two-loop approximation we find that all $P_4$ in Eq. \eqref{eqK4P4} are zero except the following ones
 \begin{gather}
  P_4^{\alpha_1\alpha_2\alpha_3\alpha_4}(i\varepsilon_{n_1},i\varepsilon_{n_2},i\varepsilon_{n_3},i\varepsilon_{n_4}) 
 =P_4^{\alpha_1\alpha_2\alpha_3\alpha_4}(i\varepsilon_{n_2},i\varepsilon_{n_1},i\varepsilon_{n_3},i\varepsilon_{n_4}) 
  =P_4^{\alpha_1\alpha_2\alpha_3\alpha_4}(i\varepsilon_{n_1},i\varepsilon_{n_2},i\varepsilon_{n_4},i\varepsilon_{n_3}) \notag \\ 
=P_4^{\alpha_1\alpha_2\alpha_3\alpha_4}(i\varepsilon_{n_2},i\varepsilon_{n_1},i\varepsilon_{n_4},i\varepsilon_{n_3}) 
=12
\left ( \frac{64}{g} \right )^2 \int_{qp} \mathcal{D}_q(i\Omega_{12}^\varepsilon) \mathcal{D}_p(i\Omega_{34}^\varepsilon)
 \end{gather}
Hence,
\begin{equation}
K_4 = \rho_0^4 \frac{12 t^2 h^{2\epsilon}}{\epsilon^2} +O(1) .
\end{equation}
By using Eq. \eqref{eqmqe3}, we obtain Eq. \eqref{eqmqe}. 

\end{widetext}



\begin{thebibliography}{99}



\bibitem{AL50} {\it 50 years of Anderson localization}, ed. by E.~Abrahams
(World Scientific, 2010).

\bibitem{Evers08}
F.~Evers and A.D.~Mirlin,  Rev. Mod. Phys. {\bf 80},   1355 (2008).

\bibitem{semicond} see, in particular, H. Stupp, M. Hornung, M. Lakner, O.
Madel, and H. v. L\"ohneysen Phys. Rev. Lett. {\bf 71}, 2634 (1993);
S. Bogdanovich and M. P. Sarachik, and R.N. Bhatt,
Phys. Rev. Lett. {\bf 82}, 137 (1999);
S. Waffenschmidt, C. Pfleiderer, and H. v. L\"ohneysen,
Phys. Rev. Lett. {\bf 83}, 3005 (1999) (3D metal-insulator transition);
W. Li, C. L. Vicente, J. S. Xia, W. Pan, D. C. Tsui,
L. N. Pfeiffer, and K. W. West,
Phys. Rev. Lett. {\bf 102}, 216801 (2009) (quantum Hall transition)
and references therein. See also review
of experimental activity in Refs.~[\onlinecite{Evers08,finkelstein90,belitz94}].

\bibitem{graphene} A.H.~Castro Neto, F.~Guinea,
N.M.R.~Peres, K.S.~Novoselov, and A.K.~Geim, Rev. Mod. Phys. {\bf 81}, 109
(2009).

\bibitem{topins} M.Z.~Hasan and C.L.~Kane, Rev. Mod. Phys.
{\bf 82}, 3045 (2010); X.-L.~Qi and S.-C.~Zhang, Rev.~Mod.~Phys. {\bf 83}, 1057
(2011).

\bibitem{Wegner} D. H\"of, F. Wegner, Nucl. Phys. B {\bf 275}, 561
  (1986); F. Wegner, Nucl. Phys. B {\bf 280}, 193
  (1987); Nucl. Phys. B {\bf 280}, 210 (1987).

\bibitem{gruzberg13} I. A. Gruzberg, A. D. Mirlin, and M. R. Zirnbauer,
Phys. Rev. B {\bf 87}, 125144 (2013).

\bibitem{HW} D-H. Lee and Z. Wang, Phys. Rev. Lett. {\bf 76}, 4014 (1996).

\bibitem{WFGC} Z. Wang, M. P. A. Fisher, S. M. Girvin, and J. T. Chalker, Phys.
Rev. B {\bf 61}, 8326 (2000).

\bibitem{BBEGM} I. S. Burmistrov, S. Bera, F. Evers, I. V. Gornyi, and A. D. Mirlin,
Ann. Phys. {\bf 326}, 1457 (2011).

\bibitem{feigelman07}
M.V.~Feigelman, L.B.~Ioffe, V.E.~Kravtsov, and E.A.~Yuzbashyan, Phys. Rev.
Lett. {\bf 98}, 027001 (2007);
M.V.~Feigelman, L.B.~Ioffe, V.E.~Kravtsov, and E.~Cuevas,
Annals of Physics {\bf 325}, 1368 (2010).

\bibitem{burmistrov12}
I.S. Burmistrov, I. V. Gornyi, and A. D. Mirlin,
Phys. Rev. Lett. {\bf 108}, 017002 (2012).

\bibitem{dellanna} L. Dell'Anna, Phys. Rev. B {\bf 88}, 195139 (2013)

\bibitem{foster12}
M.S. Foster and E.A. Yuzbashyan, Phys. Rev. Lett. {\bf 109}, 246801
(2012).

\bibitem{foster14} M. S. Foster, H.-Y. Xie, Y.-Z. Chou, Phys. Rev. B {\bf 89}, 155140 (2014).

\bibitem{finkelstein90}  A.M.~Finkelstein,  Sov. Sci. Rev. A Phys. {\bf 14}, 1
(1990).

\bibitem{belitz94} D.~Belitz and T.R.~Kirkpatrick,
  Rev. Mod. Phys. {\bf 66}, 261 (1994).

\bibitem{finkelstein10} A.M. Finkelstein, Int. J. Mod. Phys. B {\bf 24}, 1855 (2010). 

\bibitem{AA1979AAL1980} B.L. Altshuler and A.G. Aronov, Sov. Phys. JETP {\bf 50}, 968 (1979);
B.L. Altshuler, A.G. Aronov, and P.A. Lee, Phys. Rev. Lett. {\bf 44}, 1288 (1980);
B.L. Altshuler and A.G. Aronov, in
{\it Electron-Electron Interactions in Disordered Conductors}, ed.
A.J. Efros and M. Pollack, Elsevier Science Publishers,
North-Holland, 1985.


\bibitem{ES75SE84} A.L.~Efros and B.I.~Shklovskii, J. Phys. C \textbf{8}, L49 (1975);
B.I.~Shklovskii and A.L.~Efros, \textit{Electronic Properties of Doped Semiconductors},
(Springer, New York, 1984).

\bibitem{Fin198384} A. M. Finkelstein, JETP Lett. {\bf 37}, 517 (1983); Sov. Phys. JETP {\bf 53}, 97 (1983);  Sov. Phys. JETP {\bf 59}, 212 (1984).


\bibitem{Castellani1984} C.\
Castellani, C.\ DiCastro, P.A. Lee and M.\ Ma, Phys.
Rev. B \textbf{30}, 527 (1984).

\bibitem{nazarov89} Yu. V. Nazarov, Sov. Phys. JETP {\bf 68}, 561 (1989).

\bibitem{levitov97} L. S. Levitov, A. V. Shytov, JETP Lett. {\bf 66}, 214
(1997).

\bibitem{kamenev99} A. Kamenev and A. Andreev,
Phys. Rev. B {\bf 60}, 2218 (1999).

\bibitem{richardella10} A. Richardella, P. Roushan, S. Mack, B. Zhou, D.A.
Huse, D.D. Awshalom, and A. Yazdani, Science {\bf 327}, 665 (2010).

\bibitem{morgenstern02} M. Morgenstern, D. Haude, J. Klijn, and R. Wiesendanger,
Phys. Rev. B {\bf 66}, 121102(R) (2002).

\bibitem{morgenstern-2D} K. Hashimoto, C. Sohrmann, J. Wiebe,
T. Inaoka, F. Meier, Y. Hirayama, R.A. R\"omer, R. Wiesendanger, and M.
Morgenstern, Phys. Rev. Lett. {\bf 101}, 256802 (2008); S. Becker, C. Karrasch,
T. Mashoff, M. Pratzer, M. Liebmann, V. Meden, and M. Morgenstern, Phys. Rev.
Lett. {\bf 106}, 156805 (2011); M. Morgenstern, Phys. Stat. Sol. {\bf 248}, 2423
(2011);  M. Morgenstern, A. Georgi, S. Stra{\ss}er, C.R.
Ast, S. Becker, and M. Liebmann, Physica E {\bf 44}, 1795 (2012).

\bibitem{sacepe08}
B. Sacepe, C. Chapelier, T. I. Baturina, V. M. Vinokur, M. R. Baklanov, and M.
Sanquer, Phys. Rev. Lett. {\bf 101}, 157006 (2008).

\bibitem{slevin} Y. Harashima and K. Slevin, Int. J. Mod. Phys. Conf. Ser. {\bf 11}, 90
(2012); Phys. Rev. B {\bf 89}, 205108 (2014).

\bibitem{kravtsov14} M. Amini, V. E. Kravtsov, and M. Mueller, New J. Phys. {\bf 16}, 015022 (2014).


\bibitem{BGM2013} I.S. Burmistrov, I.V. Gornyi, A.D. Mirlin, Phys. Rev. Lett. {\bf 111}, 066601 (2013).

\bibitem{punnoose05} A.~Punnoose and A.M.~Finkelstein, Science {\bf 310}, 289 (2005).




\bibitem{baranov99} M.A.~Baranov, A.M.M.~Pruisken, and B.~\v{S}kori\'{c}, Phys. Rev. B {\bf 60}, 16821 (1999).


\bibitem{Wegner79} F.~Wegner, Z. Physik B {\bf 36}, 209 (1980).


\bibitem{BGM2014} I.S. Burmistrov, I.V. Gornyi, A.D. Mirlin, Phys. Rev. B {\bf 89}, 035430 (2014).

\bibitem{baranov02} M.A.~Baranov, I.S.~Burmistrov, and A.M.M.~Pruisken, Phys. Rev. B {\bf 66}, 075317 (2002).

\bibitem{HikamiWegnerB} S. Hikami, Nucl. Phys. B{\bf 215}, 555 (1983); W. Bernreuther and F.J. Wegner, Phys. Rev. Lett. {\bf 57}, 1383 (1986); F. Wegner, Nucl. Phys. B{\bf 316}, 663 (1989).

\bibitem{burmistrov07} I.S. Burmistrov and A.M.M. Pruisken, Ann. Phys. (N.Y.) {\bf 322}, 1265 (2007).

\bibitem{vortices}  E. J. K\"onig, P. M. Ostrovsky, I. V. Protopopov, and A. D. Mirlin, Phys. Rev. B {\bf 85}, 195130 (2012); L.~Fu and C.L.~Kane, Phys. Rev. Lett. {\bf 109}, 246605 (2012).

\bibitem{OGM2007} P.M. Ostrovsky, I.V. Gornyi and A.D. Mirlin, Phys. Rev. Lett. {\bf 98}, 256801 (2007); S. Ryu, C. Mudry, H. Obuse, and A. Furusaki, Phys. Rev. Lett. {\bf 99}, 116601 (2007).

\bibitem{OGM2010} P.M. Ostrovsky, I.V. Gornyi and A.D. Mirlin, Phys. Rev. Lett. {\bf 105}, 036803 (2010).

\bibitem{SM} J.H. Bardarson, J. Tworzydlo, P.W. Brouwer, and C.W.J. Beenakker, Phys. Rev. Lett. {\bf 99}, 106801 (2008);  
K. Nomura, M. Koshino, and S. Ryu, Phys. Rev. Lett. {\bf 99}, 146806 (2007).

\bibitem{ringel12} Z. Ringel, Y. E. Kraus, and A. Stern, Phys. Rev. B {\bf 86}, 045102  (2012). 


\bibitem{jia08} X.~Jia, A.R.~Subramaniam, I.A.~Gruzberg, and
S.~Chakravarty, Phys. Rev. B {\bf 77}, 014208 (2008).

\bibitem{obuse10} H.~Obuse, A.R.~Subramaniam, A.~Furusaki, I.A.~Gruzberg, and
A.W.W.~Ludwig, Phys. Rev. B {\bf 82}, 035309 (2010).

\bibitem{gruzberg11} I.A.~Gruzberg, A.W.W.~Ludwig, A.D.~Mirlin, and
M.R.~Zirnbauer, Phys. Rev. Lett. {\bf 107}, 086403 (2011).

\bibitem{Future} I.S. Burmistrov, I.V. Gornyi, A.D. Mirlin, in preparation.

\end{thebibliography}
\end{document}